\documentclass[11pt,preprint]{aastex61}

\newcommand       \AV        {A_{\rm V}}
\newcommand       \AKs       {A_{\rm K_S}}

\newcommand       \AJAKs     {A_{\rm J}/A_{\rm K_S}}
\newcommand       \CJH       {C_{\rm JH}}
\newcommand       \CJKs      {C_{\rm JK_S}}

\newcommand       \Teff      {T_{\rm {eff}}}
\newcommand       \logg      {{\rm {log}}~g}
\newcommand       \feh       {\rm{[Fe/H]}}
\newcommand       \EJHJKs    {E_{\rm JH}/E_{\rm JK_S}}
\newcommand       \EJH       {E_{\rm JH}}
\newcommand       \EJKs      {E_{\rm JK_S}}
\newcommand       \Msun      {\rm M_{\sun}}
\newcommand       \Ks        {K_{\rm S}}
\newcommand       \MKs       {M_{\rm K_{S}}}
\newcommand       \Rin       {R_{\rm in}}
\newcommand       \Rout      {R_{\rm out}}
\newcommand       \Mdust     {M_{\rm dust}}
\newcommand       \Ffil      {\rm F_{fil}}

\accepted{by ApJ February 1, 2018}

\begin{document}

\title{The Distance and the near-IR extinction \\
         of the Monoceros Supernova Remnant}

\correspondingauthor{Biwei Jiang}
\email{bjiang@bnu.edu.cn}

\author{He Zhao}
\affiliation{Beijing Normal University}
\email{hezhao@mail.bnu.edu.cn}

\author{Biwei Jiang}
\affiliation{Beijing Normal University}

\author{Shuang Gao}
\affiliation{Beijing Normal University}
\email{sgao@bnu.edu.cn}

\author{Jun Li}
\affiliation{Beijing Normal University}
\email{lijun@mail.bnu.edu.cn}

\author{Mingxu Sun}
\affiliation{Beijing Normal University}
\email{mxsun@mail.bnu.edu.cn}

\begin{abstract}

Supernova remnants (SNRs) embody the information of the influence on dust properties 
by the supernova explosion. Based on the color indexes from the 2MASS photometric 
survey and the stellar parameters from the SDSS$-$DR12$/$APOGEE and LAMOST$-$DR2$/$LEGUE 
spectroscopic surveys, the near-infrared extinction law and the distance of the 
Monoceros SNR are derived together with its nearby two nebulas -- the Rosette 
Nebula and NGC 2264. The distance is found at the position of the sharp increase 
of interstellar extinction with distance and the nebular extinction is calculated 
by subtracting the foreground interstellar extinction. The distance of the Monoceros 
SNR is determined to be 1.98\,kpc, larger than previous values. Meanwhile, the 
distance of the Rosette Nebula is 1.55\,kpc, generally consistent with previous 
work. The distance between these two nebulas suggests no interaction between them. 
The distance of NGC 2264, 1.20\,kpc, exceeds previous values. The color excess ratio, 
$\EJHJKs$, is 0.657 for the Monoceros SNR, consistent with the average value 0.652 
for the Milky Way \citep{xue16}. The consistency is resulted from that the SNR 
material is dominated by interstellar dust rather than the supernova ejecta. 
$\EJHJKs$ equals to 0.658 for the Rosette Nebula, further proving the universality 
of the near-infrared extinction law. 

\end{abstract}

\keywords{ISM: supernova remnants - dust, extinction - infrared: ISM - stars: distances}

\section{Introduction} \label{intro}

A leading school of the origin of interstellar dust (ISD) is the envelopes of 
low mass stars during their asymptotic giant branch (AGB) stage. But 
with the discovery of large amount of dust in the galaxies at high red shifts 
\citep{mai04,wat15} and in the Galactic \citep{gom12,OB15,bis16,loo17} and 
extra-galactic supernova remnants (SNRs) such as SN\,1987A \citep{mat11,ind14,
wes15,BB16} and others \citep{tem17,boc16,bev17}, supernovae (SNe) are thought 
to be more important than before in alleviating the dust budgetary problem (e.g., 
\citealp{mat09, dun11}). Dust formed in the explosive ejecta of SNe disperses 
into the interstellar medium (ISM) in the phase of SNR. Theoretical computation 
demonstrates that small grains may be completely destroyed by reverse shock, but 
very large grains can be survival and dispersed into ISM without significantly 
decreasing their sizes \citep{noz07}. The amount of dust formed by SNe 
is largely a topic of much debate. Considering all estimations in the literatures, 
the dust mass of Cassiopeia\,A (Cas\,A) SNR has an uncertainty of two orders of 
magnitude, from $\sim$$10^{-3}$\,$\Msun$ \citep{hin04} to $\sim$$0.5$\,$\Msun$ 
\citep{loo17}. For recent works, \citet{bar10} derived a cool ($\sim$35\,K) dust 
component with a mass of 0.0075\,$\Msun$. \citet{are14} found $\la$0.1\,$\Msun$ 
cold dust in the unshocked ejecta. \citet{loo17} also identified a concentration 
of cold dust in the unshocked region and derived a mass of $0.3-0.5$\,$\Msun$ of 
silicate grains, with a lower limit of $\ge$0.1\,$\Msun$. Although these values 
are in better agreement because of more sophisticated techniques and better data, 
the estimation of dust mass is still a difficult job. Because the majority of dust 
in SNRs is cold and thus radiating weakly in the far infrared (FIR), its radiation 
can hardly be detected. It is therefore hard to estimate the mass of dust produced 
by SNe when one is only detecting warm dust that makes up just a small fraction 
of the total dust \citep[usually two orders lower than the cold component,][]{gom12,
loo17}. \citet{BB16} present an alternative method. They study the late-time optical
and near-infrared (NIR) line profiles of SNRs, which will exhibit a red-blue asymmetry 
as a result of greater extinction produced by the internal dust. \citet{bev17} 
applied this approach to estimate dust mass for three SNRs, and gave an estimate 
of $\sim$$1.1$\,$\Msun$ for Cas\,A. The technique we adopt in this paper also 
exploits the extinction effects of dust rather than its infrared emission in order 
to trace all of the dust (both warm and cold components). Our approach is based 
on the fundamental principle that absolute extinction is proportional to dust mass.

The Monoceros Nebula (G205.5 +0.5) is an old \citep[$1.5\times10^5$\,yr;][]{gra82}
nebulous object that is firstly verified to be a SNR by the fine filamentary
structure observed in the Palomar Sky Atlas red plates and the non-thermal radio
emission at 237\,MHz and 1415\,MHz \citep{dav63}. It lies between the Rosette
Nebula (southeast) and NGC 2264 (north). It has the largest angular diameter,
220$\arcmin$, among the Galactic SNRs \citep{gre14}. Table \ref{targets} presents
its position, lying almost in the midplane of the Milky Way, together with that
of the Rosette Nebula and NGC 2264 that are slightly above the Galactic plane.

The Monoceros SNR has been observed in almost complete wavebands, from gamma-ray
to radio. With the observation of FERMI/LAT, \citet{kat16} suggest that the
gamma-ray emission from the Monoceros SNR is dominated by the decay of $\pi^{0}$
produced by the interaction of shock-accelerated protons with the ambient matter.
\citet{lea86} finds that the X-ray bright regions correlate well with the bright
optical filaments, but none of his six point sources seems to be a neutron star.
In optics, it appears that there are two distinct parts: one is diffuse in the
center and the other is a filamentary structure along the edge of the remnant
\citep{dav78}. Based on the observations at 60\,$\micron$, 6\,cm, 11\,cm, and
21\,cm, a new southern shell branch and a western strong regular magnetic field
are found in the region of Monoceros \citep{xiao12}.

\begin{figure}[!htbp]
\centering
\includegraphics[scale=0.9]{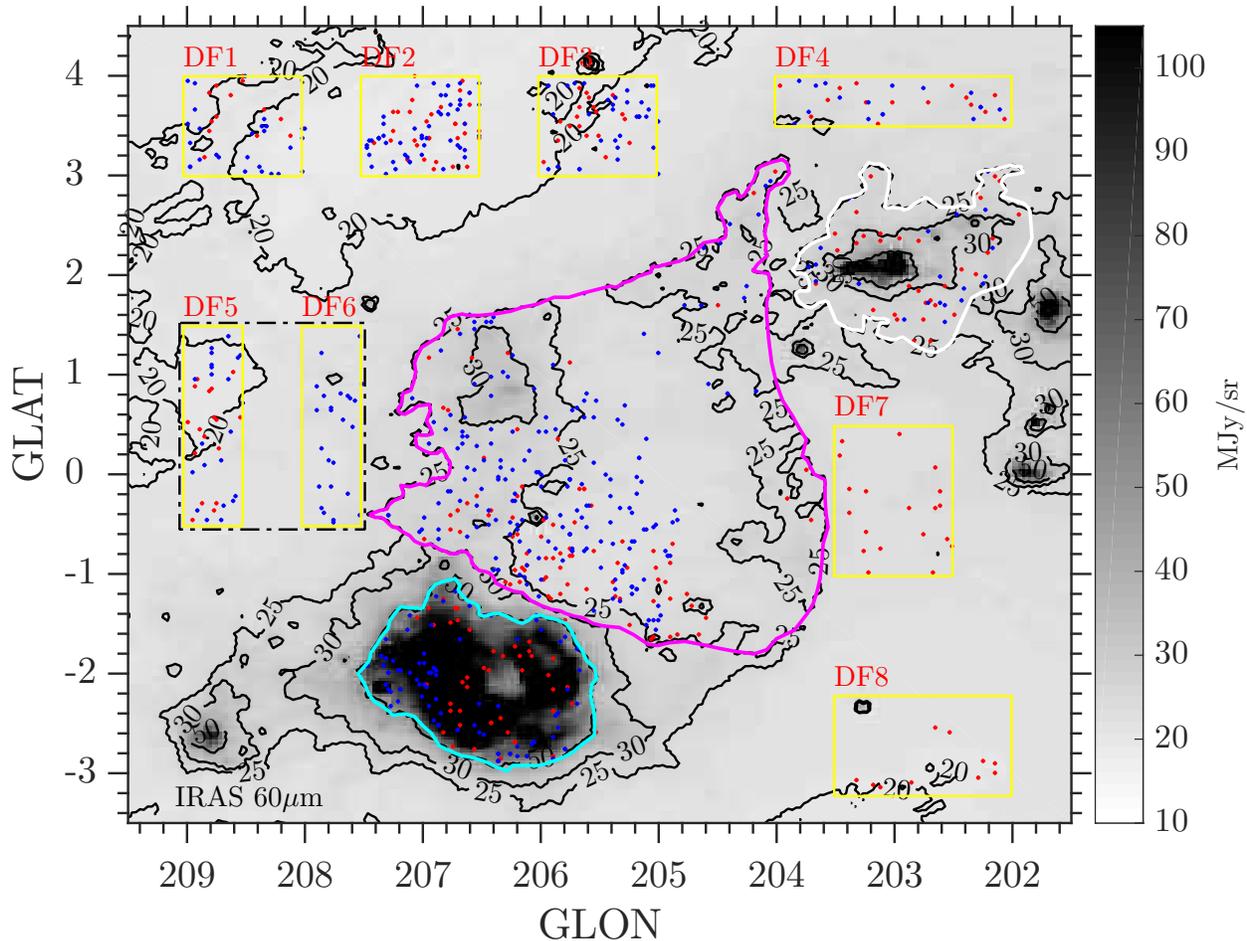}
\caption{The contours of the IRAS 60\,$\micron$ image of the target region centered 
on (Gal: $205\fdg5,+0\fdg5$). The ISM background emission in 
the Galactic plane is about $20-25$\,MJy/sr. The emission from dust in Monoceros 
SNR is $25-40$\,MJy/sr, apparently higher than adjacent ISM. Meanwhile the emission
from compact regions of the Rosette Nebula and NGC 2264 are both over 50\,MJy/sr,
with the maximum reaching about 250\,MJy/sr. The blue and red dots are the tracers 
(see Section \ref{monarea}) -- dwarfs and giants in the selected regions, respectively. 
The magenta, cyan and white lines are the borders of Monoceros SNR, Rosette 
Nebula and NGC 2264, respectively. The yellow lines are the borders of eight diffuse 
regions. Additionally, the black dot dash lines, surrounding DF5 and DF6, 
enclose the reference region mentioned in Section \ref{foreext}.
\label{IRAS}}
\end{figure}

Near the Monoceros SNR, the Rosette Nebula is a large H\,\uppercase\expandafter{\romannumeral2}
region located near a giant molecular cloud, associated with the open cluster
NGC 2244. It appears that the Rosette Nebula is overlapped with the filamentary
structure of Monoceros in the southeast \citep{dav63}. North of the Monoceros SNR,
NGC 2264 contains two astronomical objects: the Cone Nebula, an H\,\uppercase\expandafter{\romannumeral2}
region located in the southern part, and the northern part named Christmas Tree
Cluster. The Cone's shape comes from a dark absorption nebula consisting of cold
molecular hydrogen and dust. The region of Cone Nebula and the cluster is very
small (about 20$\arcmin$ in diameter), but there seems to be a much larger dust
cloud surrounding them. The rim of the cloud extends southward to the edge of
Monoceros. This is supported by \citet{dav78} and the observation of IRAS
\citep[Infrared Astronomical Satellite;][]{neu84,whe94} at 60\,$\micron$
(Figure \ref{IRAS}).

The distances of the three nebulas are not certainly determined. By making use
of the empirical surface brightness - diameter relation (the $\Sigma - D$
relation) \citep{mil74}, \citet{dav78} estimates a distance of Monoceros as
$1.6 \pm 0.3$\,kpc. Other studies result in 1.5\,kpc \citep{lea86} and 1.6\,kpc
\citep{gra82} (with the same $\Sigma - D$ relation, but different values of
parameters). For the two neighbouring nebulas, the distance of NGC 2264 is around
0.8\,kpc, as determined to be 0.715\,kpc \citep{BF63}, 0.8\,kpc \citep{wal65},
and 0.95\,kpc \citep{mor65}. For Rosette Nebula, the results are highly dispersive,
1.66\,kpc \citep{joh62}, 1.7\,kpc \citep{mor65}, and 2.2\,kpc \citep{BF63}. From
the measurement of H$\alpha$, \citet{dav78} presents a systematic change of
heliocentric radial velocities ($\rm V_{HEL}$) from north to south, which gives
some clues concerning their relative distances, and suggests that there may be
interaction between Monoceros and Rosette, while NGC 2264 is in front of them.
\citet{xiao12} also suggests that Monoceros has probably triggered part of the
star formation in the Rosette Nebula.

In this work, we try to determine both the extinction and the distance simultaneously
of the Monoceros SNR by measuring the corresponding parameters of a number of
stars in its sightline. In the same time, the extinction and distances are determined
for the two neighbouring nebulas, Rosette Nebula and NGC 2264. Stellar extinction
will increase sharply when meeting with nebula due to the higher dust density than
the diffuse medium, therefore, the distance to the nebula can then be found from
the position where the extinction increases sharply. The main steps are
shown as following:

\begin{enumerate}
  \item We determine the relation between intrinsic color index in the 
      NIR and stellar effective temperature, and use it to calculate the NIR 
      extinction and color excess for each star.
  \item Absolute magnitudes and distances of individual stars are calculated based
      on stellar parameters and photometry by using the PARSEC model.
  \item The distances to the Monoceros SNR, as well as Rosette Nebula and NGC
      2264, are derived according to position of the sharply increased extinction
      along the line of sights.
  \item The extinction produced by the SNR and the other two nebulas is derived 
      by subtracting the foreground extinction. The color excess ratio, $\EJHJKs$
      is used to describe the NIR extinction law.
  \item A rough estimation of the dust mass in the SNR is derived from its extinction.
\end{enumerate}

In Section \ref{data}, the data sets and quality controls are described. We
determine the extinction and distance of individual star in Section \ref{stellar}.
We use these results to estimate the distances of the three nebulas in Section
\ref{nebdis}. The near-infrared extinction law is derived in Section \ref{monextlaw}.
We estimate the dust mass in the region of Monoceros SNR according to its extinction
in Section \ref{dustmass}. Finally, we summarize the results and implications of this
study in Section \ref{sum}.

\begin{table}
\begin{center}
\caption{\label{targets}The geometrical information of the three targets}
\begin{tabular}{lccccc}
\tableline \tableline
    Object    & RA    & DEC   & GLON   & GLAT  & Angular diameter  \\
              & (h:m) & (d:m) & (deg)  & (deg) & (arcmin)      \\
\tableline
    Monoceros & 6 39  & 6 30  & 205.73 & 0.21  & 220           \\
\tableline
    Rosette   & 6 34  & 5 00  & 206.47 & -1.65 & 78            \\
\tableline
    NGC 2264  & 6 41  & 9 53  & 202.95 & 2.20  & 20            \\
\tableline
\end{tabular}
\end{center}
\end{table}

\section{Data and Quality Control} \label{data}

In order to complete the task, the near-infrared photometric data is taken from
2MASS, and the stellar parameters are taken from the spectroscopic surveys --
SDSS$-$DR12$/$APOGEE and LAMOST$-$DR2$/$LEGUE.

\subsection{Data} \label{dataintro}

\subsubsection{2MASS} \label{2mass}

The \emph{Two Micron All Sky Survey} (2MASS) is an all-sky photometric survey in
the near-infrared bands $JH\Ks$ \citep{coh03}. There are over 470 million stars
in the 2MASS All-Sky Point Source Catalog \citep{cut03}.

\subsubsection{APOGEE} \label{apo}

As one of the four experiments in the Sloan Digital Sky Survey \uppercase\expandafter{\romannumeral3}
(SDSS$-$\uppercase\expandafter{\romannumeral3}), \emph{Apache Point Observatory
Galactic Evolution Experiment} (APOGEE) is a high-resolution ($R \approx 22,500
$), near-infrared ($H-$band, 1.51\,$\micron - 1.70$\,$\micron$) spectroscopic
survey with high signal-to-noise ratio (about 85\% stars with $\rm S/N>100$) of
more than 100,000 Galactic red giant stars. APOGEE measures stellar parameters,
including effective temperature $\Teff$, surface gravity $\logg$, and metallicity
$\rm [M/H]$. The most recently released APOGEE catalog we use contains 163,278
stars \citep{eis11,ala15}.

\subsubsection{LEGUE} \label{leg}

The Large Sky Area Multi-Object fiber Spectroscopic Telescope (LAMOST) is a Chinese
national scientific research facility operated by National Astronomical Observatories,
Chinese Academy of Science (NAOC). LAMOST is a reflecting Schmidt telescope with
a 5.72\,m $\times$ 4.40\,m Schmidt mirror (MA) and a 6.67\,m $\times$ 6.05\,m 
primary mirror (MB). Both MA and MB are segmented. The unique design of LAMOST 
enables it to obtain 4,000 spectra in a single exposure to a limiting magnitude 
as faint as $r = 19$ at the resolution $R = 1800$ with the wavelength 
coverge of $3700< \lambda <9100$\,\AA \citep{zhao12}. The \emph{LAMOST Experiment
for Galactic Understanding and Exploration} (LEGUE) survey is one of its two key 
projects, observing both dwarf and giant stars that are used as extinction
tracers in this work. LAMOST$-$DR2$/$LEGUE, with stellar parameters, i.e. $\Teff$, 
$\logg$, $\feh$, was released in 2015 and contained more than two million sources 
\citep{deng12,deng14}, which is the dataset we use.

\subsection{Data Quality Control} \label{reduction}

In order to determine both the extinction and distance of the Monoceros SNR, as 
well as NGC 2264 and the Rosette Nebula, dwarfs and giants are chosen as extinction 
tracers and distance indicators mainly because their intrinsic colors are 
well determined by \citet{jian17} and \citet{xue16}. The preliminary operation 
combines near-infrared photometry with stellar parameters. The data is 
collated by matching the sources from 2MASS point source catalog and LAMOST$/$DR2 
within 1\arcsec. Meanwhile, the APOGEE catalog already includes the 2MASS photometry 
since the APOGEE survey was based on 2MASS.

The data quality is controlled for a precise result. The stars are picked up only
if they have full information of photometry in all the $JH\Ks$ bands and of stellar
parameters $\Teff$, $\logg$ and $\feh$. Although APOGEE measures $\rm [M/H]$ instead
of $\feh$, \citet{mes13} points out that $\rm [M/H]$ is generally close to $\feh$.
Therefore we assume that $\rm [M/H]$ is equivalent to $\feh$. The measurements
are required to fulfill the following criteria.

\begin{enumerate}
  \item The photometric error of the $JH\Ks$ bands, $\sigma_{\rm {JH\Ks}}<0.05$\,mag.
  \item The errors of stellar parameters from LEGUE, $\sigma_{\rm \Teff}<300$\,K,
      $\sigma_{\rm \logg}<0.5$\,dex, and snr$g>30$ (signal-to-noise ratio in the
      $g-$band).
  \item The errors of stellar parameters from APOGEE, $\sigma_{\rm \Teff}<300$\,K,
      $\sigma_{\rm \logg}<0.2$\,dex, and ${\rm S/N}>100$. In addition, the velocity
      scattering of multi-epoch measurments, $\rm VSCATTER <0.3$\,km/s to exclude
      binary stars.
\end{enumerate}
The different criterion in $\logg$ for LEGUE and APOGEE is caused by the much 
higher accuracy of APOGEE than LEGUE by its much higher spectral resolution.

Furthermore, the dwarf and giant stars are chosen according to the following criteria:

\begin{enumerate}
  \item 4000\,K $<\Teff<$ 7000\,K for dwarfs because of relatively uncertain 
  parameters at both lower and higher effective temperatures for the LAMOST$/$DR2 
  catalog. 4000\,K $<\Teff<$ 5200\,K for G$-$ and K$-$type red giants for which 
  the intrinsic near-infrared colors are well determined by \citet{xue16}. Although 
  G$-$ and K$-$type giants have a $\Teff$ range extending to 3600\,K, but most 
  giants with 3500\,K $<\Teff<$ 4000\,K have $\logg <1$, i.e. they are red supergiants.	
  \item $\logg>4$ for dwarfs, and $1<\logg<3$ for giants. \citet{wor16}
  set a value of $\logg=3.5$ as the boundary of giant and dwarf. Taking the typical
  value of $\Delta \logg$ of LEGUE ($\sim 0.5$\,dex) into account, \citet{jian17}
  shifted the boundary and stars with $3<\logg<4$ are dropped to avoid ambiguity,
  which has little effect on the result thanks to the numerous stars in the database.
  \item $-0.5<\feh<0.5$ for both dwarfs and giants. Both metal-poor and metal-rich
  stars are removed to reduce the influence of metallicity on intrinsic colors
  in near-infrared bands. Moreover, this range of metallicity is much precisely
  determined.
\end{enumerate}

Under these criteria, 374,052 dwarfs and 90,741 giants (45,444 from LEGUE and
45,297 from APOGEE) are selected to consist the star sample for our study of the
relation between stellar intrinsic colors and effective temperatures.

Based on our criteria, stars fainter than $\Ks=14.4$\,mag will be excluded,
most of which are far away or highly obscured by dust. But our star sample can
still reach as far as $8\,$kpc, most within $6\,$kpc, covering the three targets
(around $2\,$kpc). Additionally, it is enough to trace the extinction of the faint
SNR. Meanwhile, such depth may be unable to trace the dense regions of the three nebulae.

\subsection{Selction of the Area of the Monoceros SNR}\label{monarea}

SNRs radiate both in radio and infrared. Since we are interested in the extinction
and dust of SNRs, the dust emission map would be the appropriate indicator of
the region of SNR. As dust dominates the infrared emission between 5\,$\micron$
and 600\,$\micron$ \citep{dra11}, we make use of the observation by IRAS at
60\,$\micron$ to trace the warm dust towards the line of sight of a $7^\circ
\times 7^\circ $ field centered at (Gal: $205\fdg5,+0\fdg5$), almost the
very center of the Monoceros SNR (Figure \ref{IRAS}). The whole field contains
2,725 stars all picked from our star sample described above, which form a sub-sample
to study the extinction and distance of stars and nenulas. We will use it to
analyze the uncertainties of the derived distance in Section \ref{Gaia}.

According to the contour map of the target regions (Figure \ref{IRAS}), 
we determine the bounds of the faint SNR by the 25\,MJy/sr contour (the 
magenta line), while by the 50\,MJy/sr contour for the compact region of Rosette 
Nebula (the cyan line). For NGC 2264, the bound is also defined by the 
25\,MJy/sr contour (the white line) in order to include as many as possible 
stars for tracing its extinction. After defining the borders of the nebulas, the 
'tracing stars' are extracted from the sub-sample in an irregular polygonal field 
for Monoceros which basically follows the bound defined by infrared flux, and so 
is done for the other two nebulas. In order to study the foreground extinction, 
we additionally select eight rectangular diffuse fields (DFs) around the three 
nebulas, where no obvious dust emission is visible. The number of selected stars 
in each field are displayed in Table \ref{starchoose}.

\begin{table*}
\begin{center}
\caption{\label{starchoose}Number of stars in each selected nebular regions and
eight diffuse fileds (DFs) (cf. Figure \ref{IRAS}).}
\begin{tabular}{cccccccccccc}
\tableline \tableline
                & Monoceros & Rosette & NGC 2264 & DF1&DF2&DF3&DF4&DF5&DF6&DF7&DF8  \\
\tableline
    Dwarf       & 194  & 97   & 23 & 28 & 53 & 34 & 17 & 23 & 25 & 0  & 0  \\
    Giant       & 85   & 47   & 33 & 13 & 33 & 20 & 15 & 17 & 0  & 19 & 10 \\
    Total       & 279  & 144  & 56 & 41 & 86 & 54 & 32 & 40 & 25 & 19 & 10 \\
\tableline
\end{tabular}
\end{center}
\end{table*}

\section{Calculation of Stellar Extinction and Distance} \label{stellar}

\subsection{Intrinsic color indexes} \label{intcolor}

We determine stellar intrinsic color indexes between band $\lambda_1$ and
$\lambda_2$, $C_{\lambda_1\lambda_2}^0$, from their $\Teff$ measured by APOGEE
or LEGUE. \citet{duc01} suggest that the stars around the blue edge in the $\Teff
- C_{\lambda_1 \lambda_2}$ diagram have the smallest extinction. For large sky
survey projects, such as LEGUE and APOGEE, extinction-free stars are included
and appear as the bluest ones in the $\Teff - C_{\lambda_1\lambda_2}$ diagram.
That is to say the observed colors of these stars are indeed their intrinsic
colors. By fitting $C_{\lambda_1\lambda_2}$ of the chosen extinction-free stars
in some temperature intervals, an analytical relation of $C_{\lambda_1 \lambda_2}^0$
with $\Teff$ can be derived. This method has recently been applied to determining
stellar intrinsic colors in infrared \citep{WJ14,xue16,jian17}. Here we adopt
the analytical function dertermined by \citet{xue16} to calculate the intrinsic
colors, $\CJH^0$ and $\CJKs^0$, for giants:

\begin{equation}\label{giantCJH0}
\rm \CJH^0 = 6.622 \times exp\left(-\frac{\Teff}{1846K}\right) + 0.019
\end{equation}

\begin{equation}\label{giantCJKs0}
\rm \CJKs^0 = 20.285 \times exp\left(-\frac{\Teff}{1214K}\right) + 0.209
\end{equation}

For the dwarf stars which are not studied in \citet{xue16}, the relation of $C_
{\lambda_1\lambda_2}^0$ with $\Teff$ is derived in the same way through the
selected LAMOST$/$LEGUE dwarfs. Firstly, the stars are binned according to their
$\Teff$ with a 50\,K step from 4000\,K to 7000\,K. Then the bluest 5\% stars are
extracted from each bin and the median value of their colors is taken as the
intrinsic color index in each bin. A quadratic function is used to fit the median
color indexes of the bluest 5\% dwarfs and $\Teff$:

\begin{equation}\label{dwarfintcolor}
\rm C^0_{\lambda_1\lambda_2}=a_0+a_1 \times \left(\frac{\rm{T_{eff}}}{1000{\rm K}}
\right)+a_2 \times \left(\frac{\rm{T_{eff}}}{1000{\rm K}}\right)^2.
\end{equation}

\begin{figure}[!htbp]
   \centering
   \includegraphics[scale=0.8]{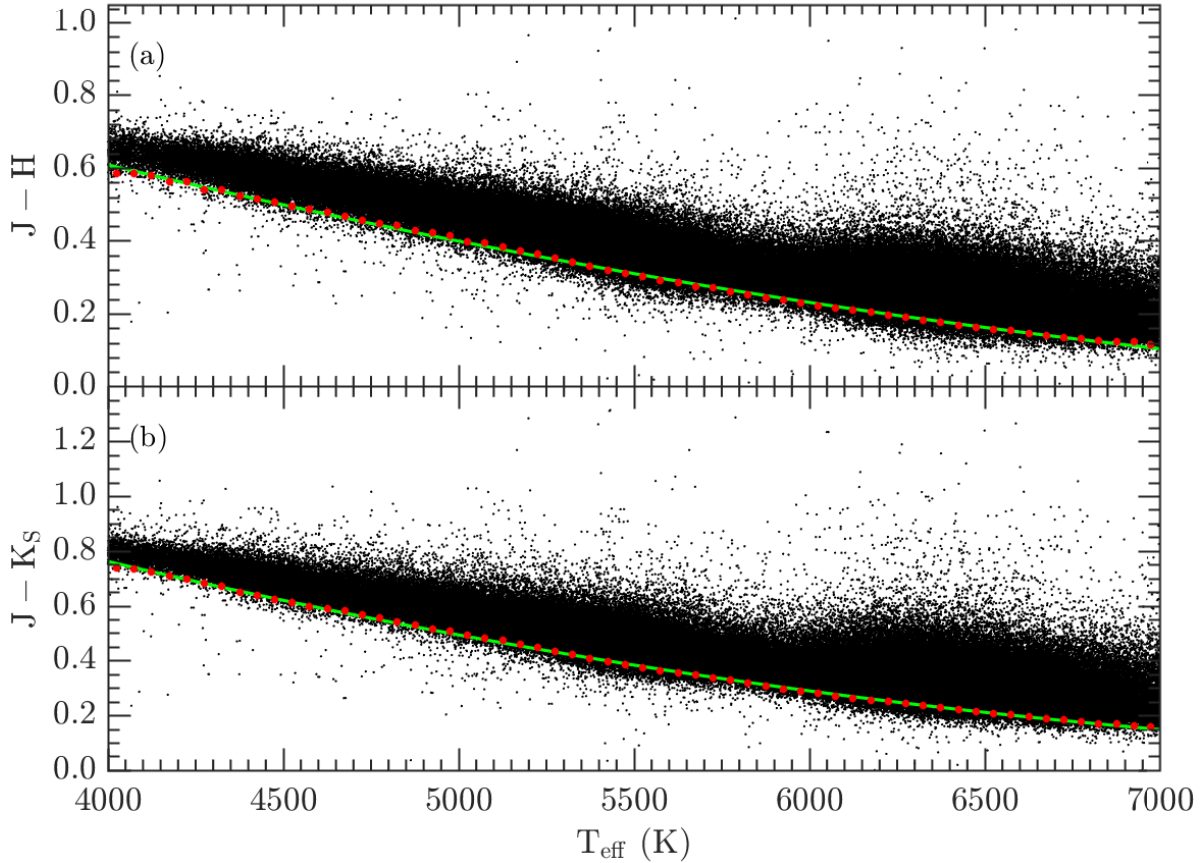}
   \caption{Determination of the relation of the intrinsic colors, $\CJH^0$ and
   $\CJKs^0$, with $\Teff$, for dwarfs. The red dots denote the median colors of
   the 5\% bluest stars in each $\Teff$ bin and the green lines are the fitting
   curves.
\label{ctdwarf}}
\end{figure}

\noindent The result is shown in Figure \ref{ctdwarf} and the coefficients for
$\CJH^0$ and $\CJKs^0$ are listed in Table \ref{intcoedwarf}. High consistency
is found with the very recent determination of intrinsic colors for dwarfs
by \citet{jian17}. The difference is no larger than 0.05 for $\CJKs^0$, 0.005
for $\CJH^0$.

As discussed by \citet{jian17}, the uncertainty of intrinsic color index comes
from a few contributors and can be expressed as:

\begin{equation}\label{intcolorerr}
\rm \sigma_{C_{J\lambda}} = \sqrt{\sigma_{para}^2 + \sigma_{\feh}^2
+ \sigma_{ratio}^2 }
\end{equation}

\noindent where $\sigma_{\rm para}$ represents the error from the uncertainties
of photometry and stellar parameters, and we finally yields 0.002 for dwarfs and
0.003 for giants by a Monte Carlo simulation. The specific technique of
the simulation is presented in detail in Section \ref{MC}. $\sigma_{\feh}$ refers
to the influence of $\feh$, we suggest an error of 0.02 for dwarfs and 0.04 for
giants, based on the discussion in Section \ref{affectfeh}.
$\rm \sigma_{ratio}$ refers to the error induced from the bluest fraction we adopt
to choose extinction-free stars. \citet{jian17} discussed different fractions and
their effect on the intrinsic colors, and set the error as 0.02.

\subsubsection{The Monte Carlo Simulation} \label{MC}

Monte Carlo simulation (MCS) is a simple way to estimate the statistical 
uncertainty caused by the stellar parameter measurement and photometry. Firstly, 
we assume a Gaussian error distribution on each observed data point in the $\Teff 
- C_{\lambda_1\lambda_2}$ plane and on the estimated errors in $JH\Ks$ magnitude 
and $\Teff$. The peak value of the distribution is the observed value of each 
parameter, like colors and $\Teff$, and the Gaussian has a width equals to the 
estimated error. Then a random data point is sampled for each observed point from 
two independent Gaussian functions because the colors and $\Teff$ are determined 
independently,

\begin{equation}
f\left(x;A,\mu,\sigma\right)=A~{\rm{exp}}\left[-\frac{\left(x-\mu\right)^2}
{2\sigma^2}\right],
\end{equation}

\noindent where $x$ is the color or $\Teff$, $\mu$ and $\sigma$ are 
correspondingly the observed values and estimated errors, respectively. We subsequently
redid the fitting described in Section \ref{intcolor} with these randomly sampled
points to get new coefficients. LAMOST$/$LEGUE dwarfs make up the sample set to
determine intrinsic colors for dwarfs. Meanwhile, for giants, we still follow the
data set and functional form of \citet{xue16}.

\begin{figure}[!htbp]
   \centering
   \includegraphics[scale=0.8]{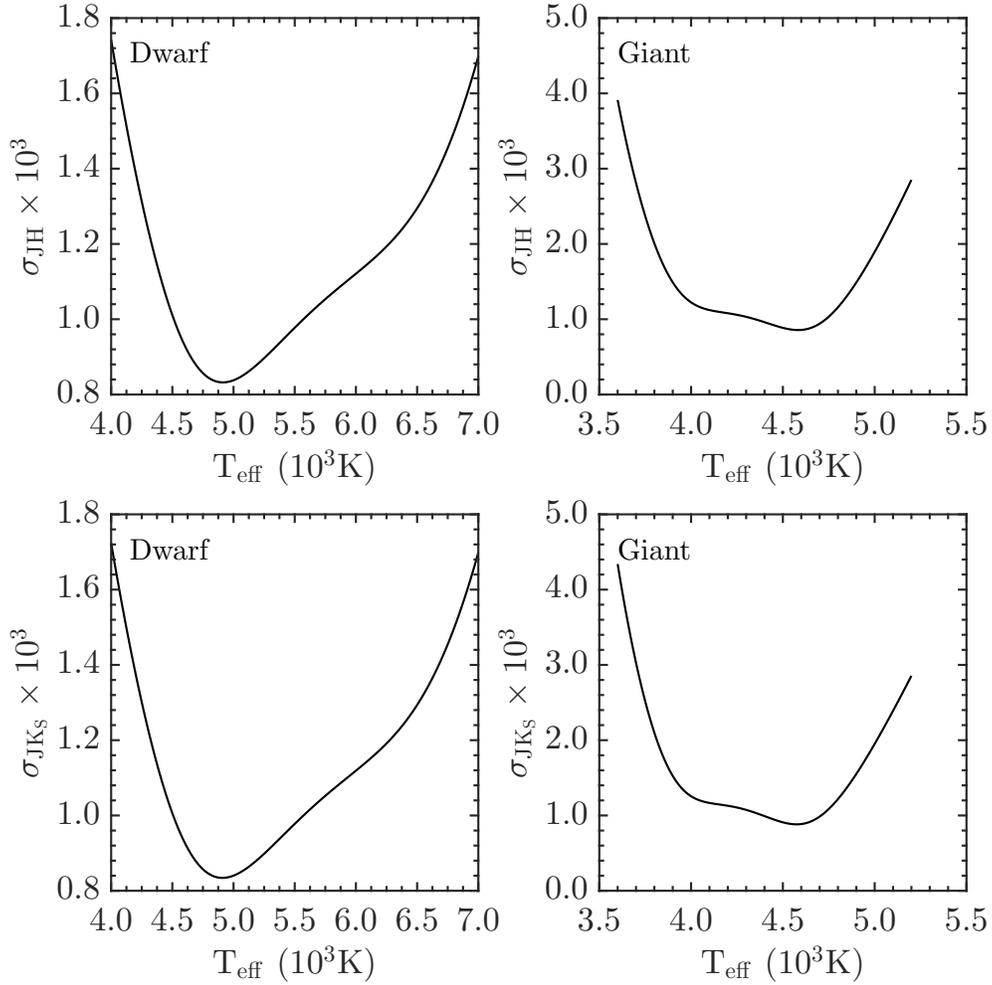}
   \caption{The uncertainties of intrinsic colors caused by the errors of photometry
   and $\Teff$, derived from the Monte Carlo simulation.
   \label{MCS}}
\end{figure}

This process is carried out 20,000 times to yield an overall distribution 
of coefficients, and the standard deviation of the distribution is the uncertainty 
of coefficients which are listed in Table \ref{intcoedwarf} and \ref{intcoegiant}.
Furthermore, the standard deviations of intrinsic colors can also be calculated 
by these sets of coefficients. As the coefficients are correlative with each other, 
we take some typical temperatures and calculate the intrinsic colors by the MCS 
result. The errors are derived from the resultant distribution and presented in 
Figure \ref{MCS}. We can find that the errors are no larger than 0.002 for dwarfs, 
and no larger than 0.003 for giants. Although $\sigma_{\Teff}$ is on the order 
of one hundred kelvin, and photometric errors are hundredth magnitude, the statistical 
method based on the large sample makes these measured and observed uncertainties 
have very weak influence on the intrinsic colors.

\begin{table*}[!htbp]
\begin{center}
\caption{\label{intcoedwarf}The fitting coefficients of intrinsic colors for
dwarfs, and their standard errors derived by Monte Carlo simulation.}
\begin{tabular}{ccccc}
\tableline \tableline
                   & $a_0$               & $a_1$                & $a_2$               \\
\tableline
$\rm C_{\rm JH}^0$ & 1.8568($\pm$0.0203) & -0.3944($\pm$0.0072) & 0.0206($\pm$0.0006) \\
$\rm C_{\rm JK}^0$ & 2.4715($\pm$0.0201) & -0.5550($\pm$0.0071) & 0.0319($\pm$0.0006) \\
\tableline
\end{tabular}
\end{center}
\end{table*}

\begin{table*}[!htbp]
\begin{center}
\caption{\label{intcoegiant}The same as Table \ref{intcoedwarf}, but for giants,
with the functional form following \citet{xue16}: $\rm C_{\lambda_1\lambda_2}^0
= a_0 \times exp\left(-\frac{\Teff}{a_1}\right) + a_2 $.}
\begin{tabular}{ccccc}
\tableline \tableline
                   & $a_0$              & $a_1$         & $a_2$             \\
\tableline
$\rm C_{\rm JH}^0$ & 6.622($\pm$1.125)  & 1846($\pm$60) & 0.019($\pm$0.021) \\
$\rm C_{\rm JK}^0$ & 20.285($\pm$2.356) & 1214($\pm$29) & 0.209($\pm$0.014) \\
\tableline
\end{tabular}
\end{center}
\end{table*}

\subsubsection{The Influence of $\feh$} \label{affectfeh}

\begin{figure}[!htbp]
   \centering
   \includegraphics[scale=0.45]{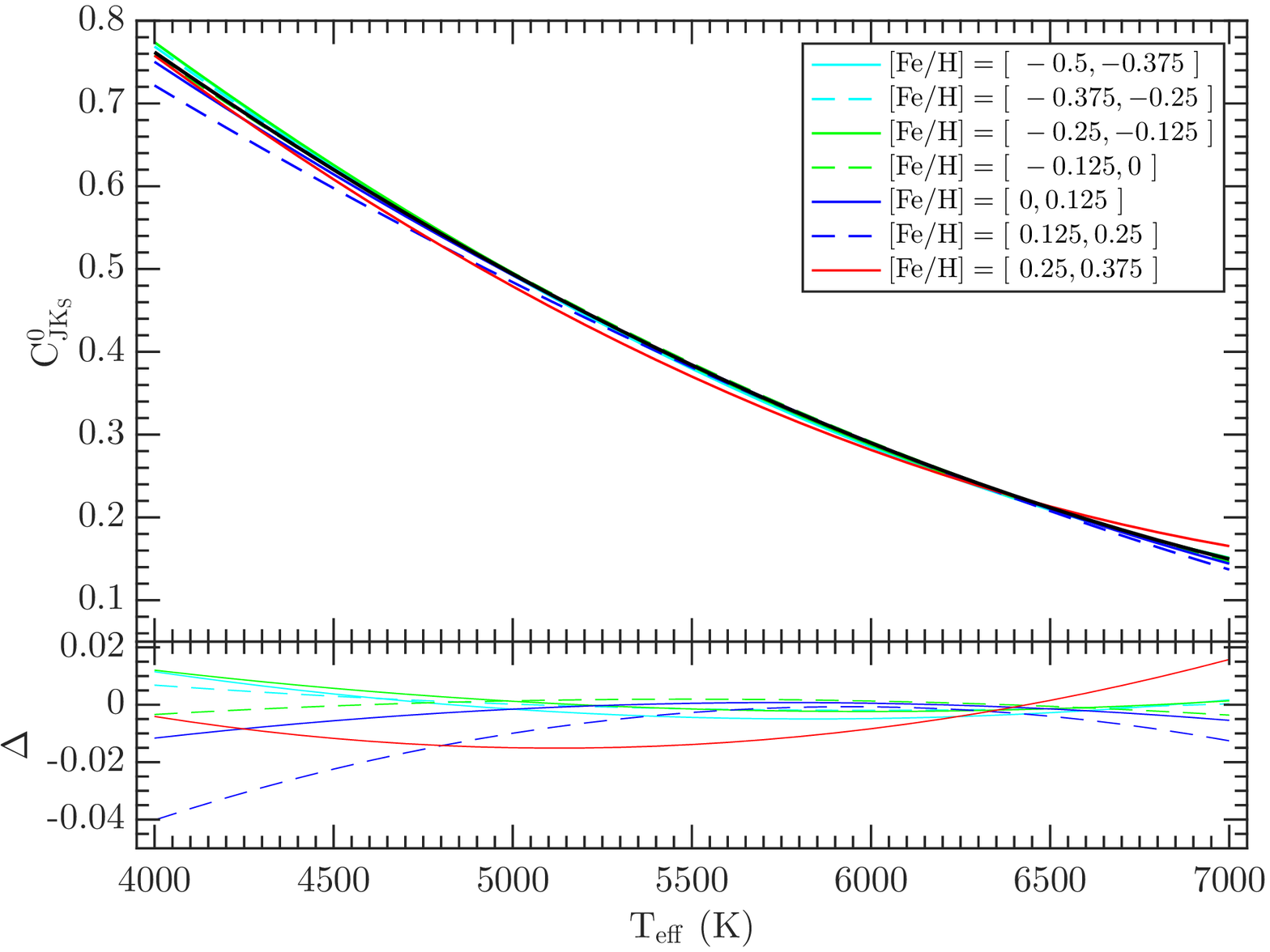}
   \includegraphics[scale=0.45]{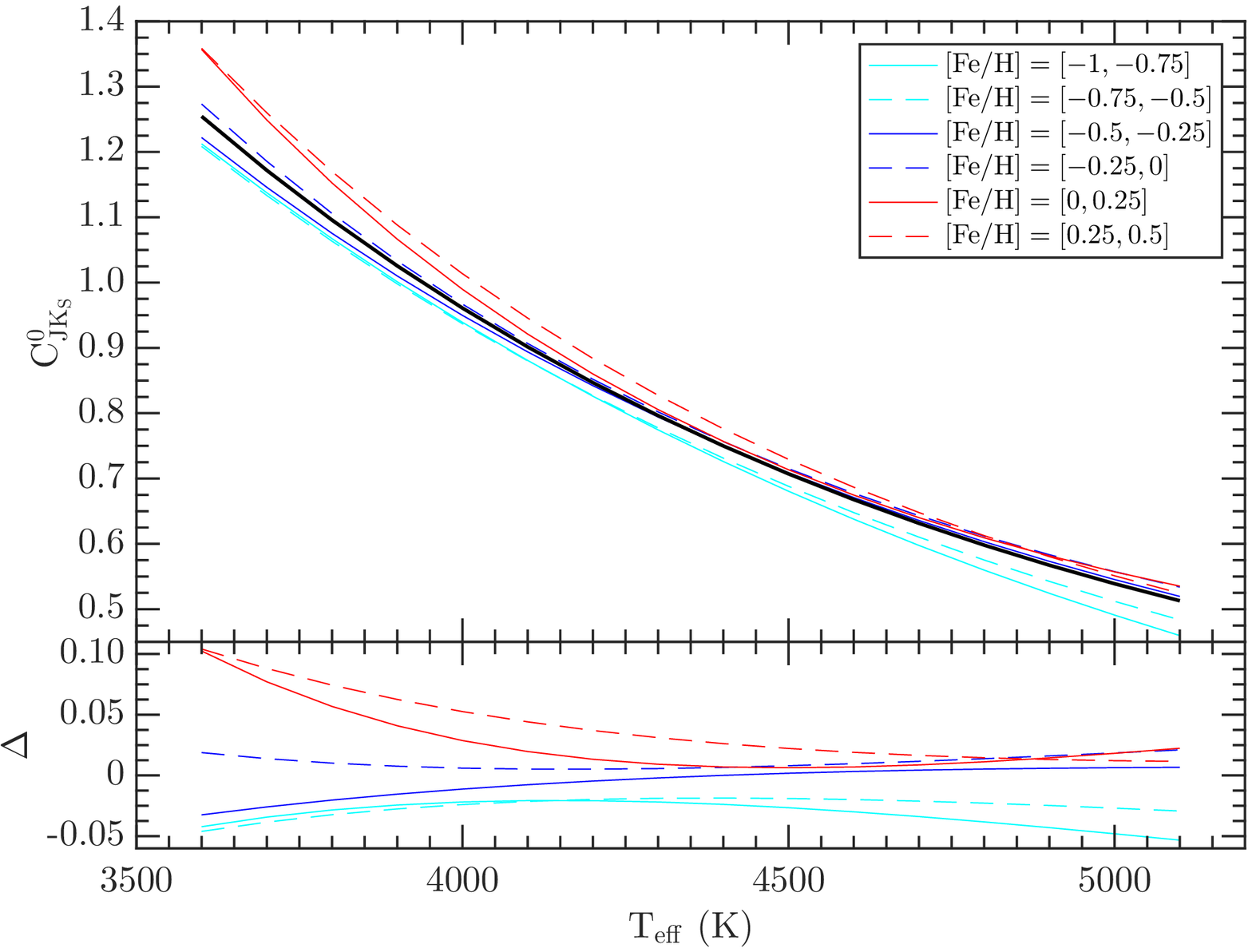}
   \caption{The influence of $\feh$ on intrinsic colors of dwarfs (left) and
   giants (right). In the upper panels, the intrinsic colors derived from different
   $\feh$ bins are present in different colors and linstyles, the black solid
   line presents the result derived using the whole $\feh$ samples. In the lower
   panels, colored lines show the differences between the corresponding bins and
   the whole samples' result.
   \label{fehcom}}
\end{figure}

\citet{jian17} analyzed the influence of [Fe/H] on the infrared intrinsic colors.
They found that the difference between metal-normal and metal-poor groups
(with a border at $\feh = -0.5$) is a few percent magnitude, no larger than 0.06.
For a higher accuracy, dwarfs are further divided into 8 groups from $\feh=-0.5$ 
to $\feh=0.5$ with a step of 0.125\,dex, and giants are divided into 6 groups
from $\feh=-1$ to $\feh=0.5$ with a step of 0.25\,dex. In each $\feh$ bin, $\CJKs^0$
is determined by the method described in Section \ref{intcolor}. Figure \ref{fehcom}
shows the fitting results and the influence of $\feh$ on the intrinsic color.

Metal-rich stars account for a pretty small proportion of both dwarfs and giants,
which leads to removing the last group of dwarfs and the abnormal
behaviours of the fitting curves (the dashed blue line and solid red line in Figure
\ref{fehcom} (left)) The differences for dwarfs are mainly within $[-0.02,~0.02]$,
so we take 0.02 as the dispersion of intrinsic colors caused by the variation of
metallicity. For giants, the differences are much larger, especially at low $\Teff$.
The dispersion rises to 0.04 for 4000\,K $<\Teff<$ 5200\,K. At low $\Teff$, the
dispersion increases for both dwarfs and giants, reaching almost 0.1\,mag
for giants when $\Teff < 4000$\,K. But it may partly come from the uncertainty of
stellar parameters at low $\Teff$ in addition to metallicity.

We prefer to taking these uncertainties (0.02 for dwarfs and 0.04 for giants) as
a part of the total uncertainty of our intrinsic color model rather than deriving
the relation between them. It's because: 1) $\Teff$ is the dominating factor for
the intrinsic colors while $\feh$ has a much weaker effect in near-infrared; 2)
The mean error of $\feh$ for dwarfs is about 0.14\,dex, which constrain the bin
box size; 3) There are not enough metal-poor and -rich stars to complete the fitting.

\subsection{$\AKs$: Interstellar Extinction in the $\Ks$ Band} \label{stellaraks}

The color excess is calculated straightforward after subtracting the intrinsic
one from the observed. The extinction in the $\Ks-$band, $\AKs$, must also be
derived in order to calculate stellar distance. The conversion from color excess,
$\EJKs$, to the extinction, $\AKs$, depends on extinction law in principle. The
near-infrared extinction law is commonly expressed by a power law $A_\lambda
\propto \lambda^{-\alpha}$. The universality of the near-infrared extinction law
\citep{WJ14} brings the convenience to convert the color excess into the absolute
extinction in the $\Ks-$band. Based on the all-sky survey data, \citet{xue16}
derived an average $\EJHJKs = 0.652$, which corresponds $\alpha = 1.79$ and $\AJAKs
= 2.72$. This conversion factor is adopted to convert $\EJKs$ to $\AKs$. The
uncertainty of $\AKs$ is then

\begin{equation}
\sigma_{\rm \AKs} = \sigma_{\rm \EJKs} / 1.72,
\end{equation}
and
\begin{equation}
\sigma_{\rm \EJKs} = \sqrt{ \sigma_{\rm J}^2 + \sigma_{\rm \Ks}^2 +
\sigma_{\rm \left(J-\Ks\right)_0}^2 },
\end{equation}

\noindent where $\sigma_{\rm \left(J-\Ks\right)_0}$ is the uncertainty of intrinsic
color discussed in Section \ref{intcolor}, and $\sigma_{\rm J}$ and $\sigma_{\rm
\Ks}$ are the observed errors.

\subsection{The Absolute Magnitude} \label{stellarmks}

We use the \emph{PAdova and TRieste Stellar Evolution Code} (PARSEC) to compute
stellar absolute magnitudes. The new PARSEC is an updating of the Padova database，
which can calculate sets of stellar evolution tracks \citep{bre12}. We
obtain stellar evolution tracks calculated by PARSEC through CMD 3.0. CMD 3.0\footnote{
CMD is being extended/updated every few months, and the last version is always 
linked in \url{http://stev.oapd.inaf.it/cmd}} is a set of routines that provide 
interpolated isochrones in a grid, together with stellar parameters and absolute 
magnitudes transformed into various photometric systems \citep[see][]{gir02, gir04}. 
The isochrone grids we use in this work have a metallicity step of 0.001\,dex 
between $0.005 < Z < 0.048$ and an age spacing of $\Delta \log(t)=0.05$\,Gyr.

For each star, we select the isochrone closest in metallicity and then the $\Ks-$band
abosolute magnitude, $\MKs$, is calculated by a two-dimensional cubic interpolation
with neighboring grid points in the corresponding $\Teff$ and $\logg$ plane, rather
than directly adopting the closest point. In this way, the accuracy of $\MKs$ is
improved in the low density area. Additionally, for a query star with a specific
type, the grid points are filtrated by the parameter `$stage$', which indicates
the stellar evolution phase, to alleviate the contamination of other type stars.
If a star lies out of the network constructed by the theoretical isochrones, the
grid points will focus on one side of it and extrapolation is needed to calculate
$\MKs$. In such case, no calculation is done for this star because errors and 
uncertainties would be unpredictable.

The typical uncertainty of $\MKs$ calculated by the PARSEC code is composed of
two parts:

\begin{equation}
\rm \sigma_{total} = \sqrt{\sigma_{para}^2  + \sigma_{inter}^2}
\end{equation}

\noindent where $\sigma_{\rm para}$ is the contribution by stellar parameters'
error, and $\sigma_{\rm inter}$ is caused by the interpolation.

\begin{figure}[!htbp]
   \centering
   \includegraphics[scale=0.9]{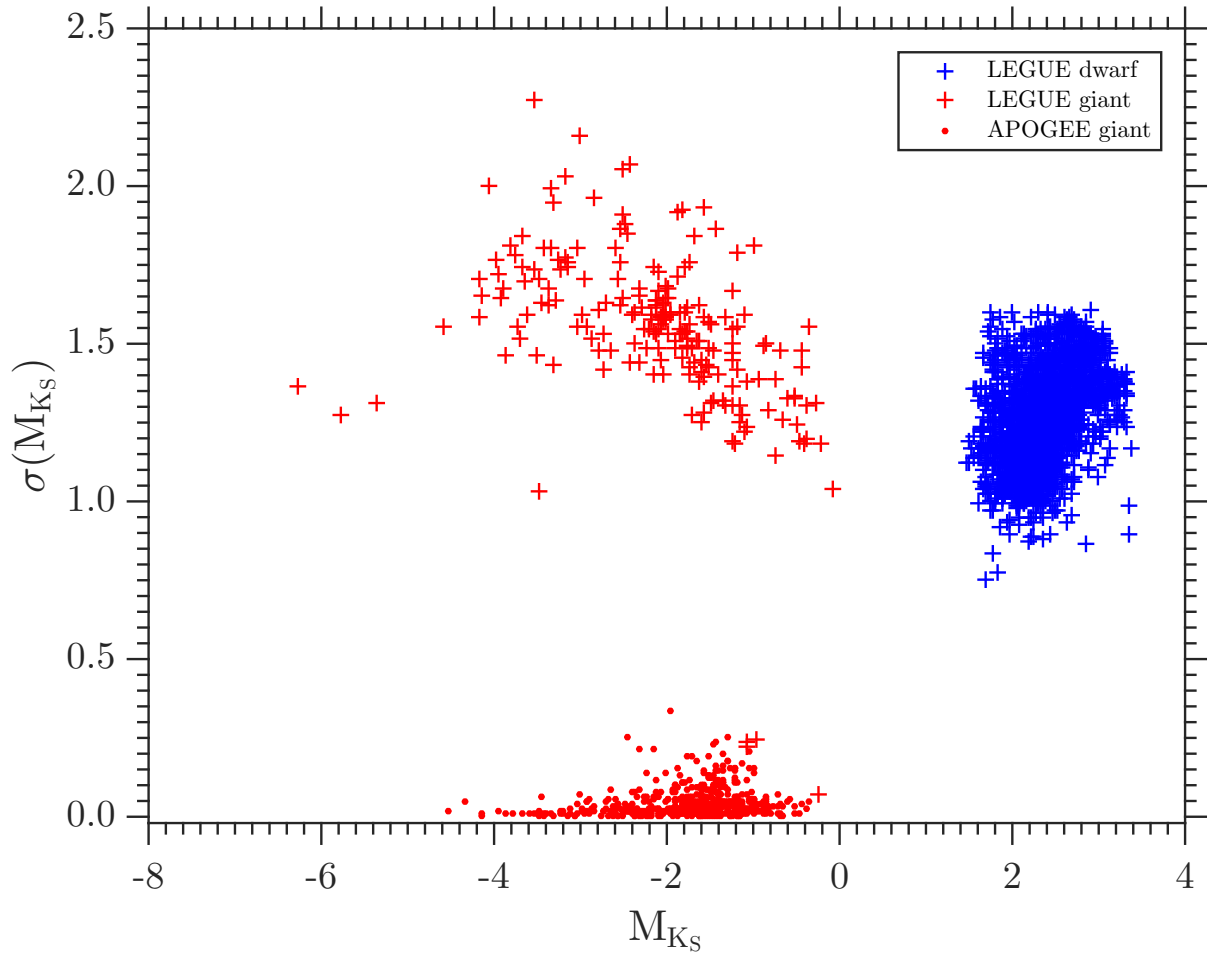}
   \caption{Uncertainty of $\MKs$ caused by the errors of stellar parameters.
   The sample stars gather into three distinct parts: blue and red crosses are
   dwarfs and giants from LEGUE, respectively, and red dots are giants from
   APOGEE.
   \label{parerr}}
\end{figure}

\citet{sch14} present a simple method to estimate $\sigma_{\rm para}$. For each
star, a new set of stellar parameters is constructed by adding the errors, i.e.
$\Teff \pm \Delta_{\rm \Teff}$, $\logg \pm \Delta_{\rm \logg}$, and
$\feh \pm \Delta_{\rm \feh}$, which is taken as a new input to calculate the
lower and upper limits of $\MKs$. Consequently, the range of $\MKs$ is calculated
and the half difference of the lower and upper limits with $\MKs$ is regarded as
the uncertainty. This method is applied to the 2,725 sub-sample stars mentioned
in Section \ref{monarea}, and $\MKs$ is successfully derived for 2,218
stars. While the remainder lie out of the theoretical network, so as mentioned
above, they are beyond calculation and dropped. Figure \ref{parerr} presents the
variation of the error $\sigma_{\MKs}$ with $\MKs$. The sample stars gather into
three distinct parts, associating with their data sources and stellar types.
There is no correlation between $\sigma_{\MKs}$ and $\MKs$, while stars observed
by LAMOST generally have significantly higer $\sigma_{\MKs}$ than APOGEE. This
can be understood by the larger error in stellar parameters of the LAMOST survey,
in particular the apparently lower quality in $\logg$ and $\feh$ than the APOGEE
survey.

For a sample star, we take the interpolated error of its closest grid
point as its $\sigma_{\rm inter}$. The interpolated error equals the difference
between intrinsic value of $\MKs$ of the grid point and the interpolated value
calculated by adjacent ones with $\Delta \Teff<$ 200\,K and $\Delta \logg<$ 0.2\,dex.
Mostly, $\sigma_{\rm inter}$ is smaller than 0.05 and negligible in comparison
with $\sigma_{\rm para}$.

The errors we discuss above do not include the contribution of the PARSEC model
itself. \citet{sch14} discussed the differences between the PARSEC isochrones and
the Basel3.1 model library \citep{lej97}. They suggest that systematic differences
exist in calculating the magnitudes and distances between the two libraries,
significant for cool, metal-poor M giants. Fortunately, we only take use of the
G$-$ and K$-$type giants, which may not be seriously affected.

\subsection{The Stellar Distance}\label{stellardis}

The distance of individual star is calculated by

\begin{equation}
\rm D(pc) = 10^{ \left[ \left( m_{K_S} + 5 - \MKs - \AKs \right) / 5 \right] },
\end{equation}

\noindent where $m_{\rm K_S}$, $\MKs$, and $\AKs$ are the apparent magnitude,
absolute magnitude and extinction magnitude in $\Ks-$band, respectively. 
According to the error analysis above, the relative uncertainty of distance is:

\begin{equation}
\rm \sigma_{D} / D = 0.46~( \sigma_{m_{K_S}} + \sigma_{\MKs} + \sigma_{\AKs} ).
\end{equation}

\begin{figure*}[!htbp]
   \centering
   \includegraphics[scale=0.9]{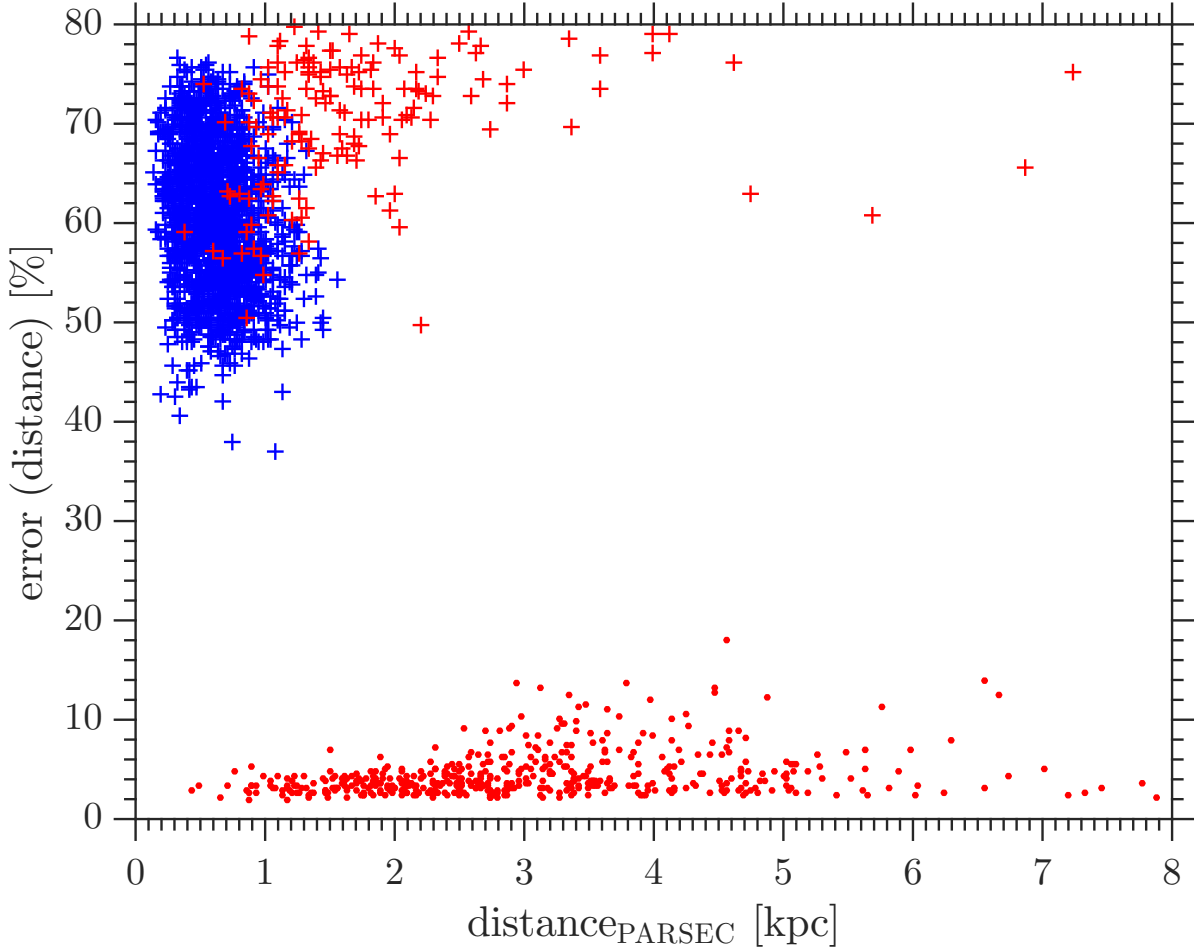}
   \caption{The relative distance error. The distance errors account the contributions
   both of the stellar parameters and the interpolation. The red dots are APOGEE
   giants, the blue and red crosses are LEGUE dwarfs and giants, respectively.
   \label{fracderr}}
\end{figure*}

For the 2,218 sample stars with $\MKs$ available, the relative error of distance
is shown in Figure \ref{fracderr}. As predictable, the errors for the LAMOST stars
(both dwarfs and giants) are significant, mostly above 50\% from the uncertainty
of derived absolute magnitude $\MKs$. On the other hand, the APOGEE giants appear with
much smaller uncertainty, mostly around $2-5$\% and never superseding 20\%, even
when the distance reaches 8\,kpc. Consequently, the LAMOST dwarfs may be problematic
in describing the run of reddening towards the targets. Meanwhile, as most dwarfs
are located within 1\,kpc, this effect is weak for the Monoceros and Rosette nebulas,
while non-negligible for the closer object NGC 2264.

\subsection{The Distance from Parallax} \label{Gaia}

Recently, the first version of data from the European Space Agency (ESA)'s $Gaia$
mission is released \citep{gaia16a,gaia16b}. It contains the $Tycho-Gaia~Astrometric
~Solution$ \citep[TGAS,][]{mic15} catalog, which provides stellar parallaxes for
about 2 million stars. The distances computed by parallaxes are independent of
stellar parameters and stellar model, which is a very good examination of the
distances derived by our method.

With the requirement of the error of parallactic distance less than 20\%, matching
TGAS with LEGUE and APOGEE results in 38,222 dwarfs and 1,468 giants (996 from
LEGUE, 472 from APOGEE). Among them, there are 143 dwarfs and 4 giants in our
target region.

\begin{figure*}[!htbp]
   \centering
   \includegraphics[scale=0.45]{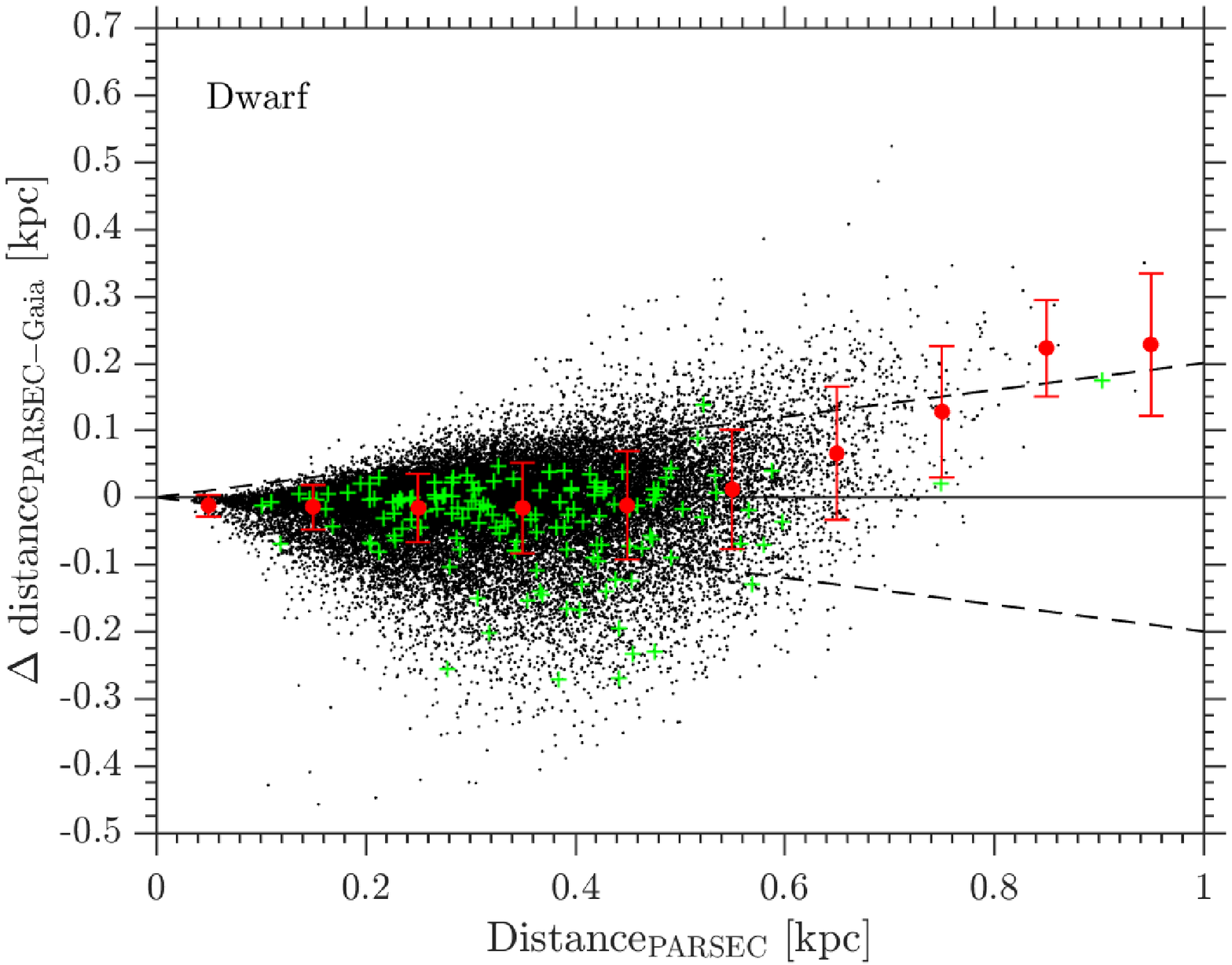}
   \includegraphics[scale=0.45]{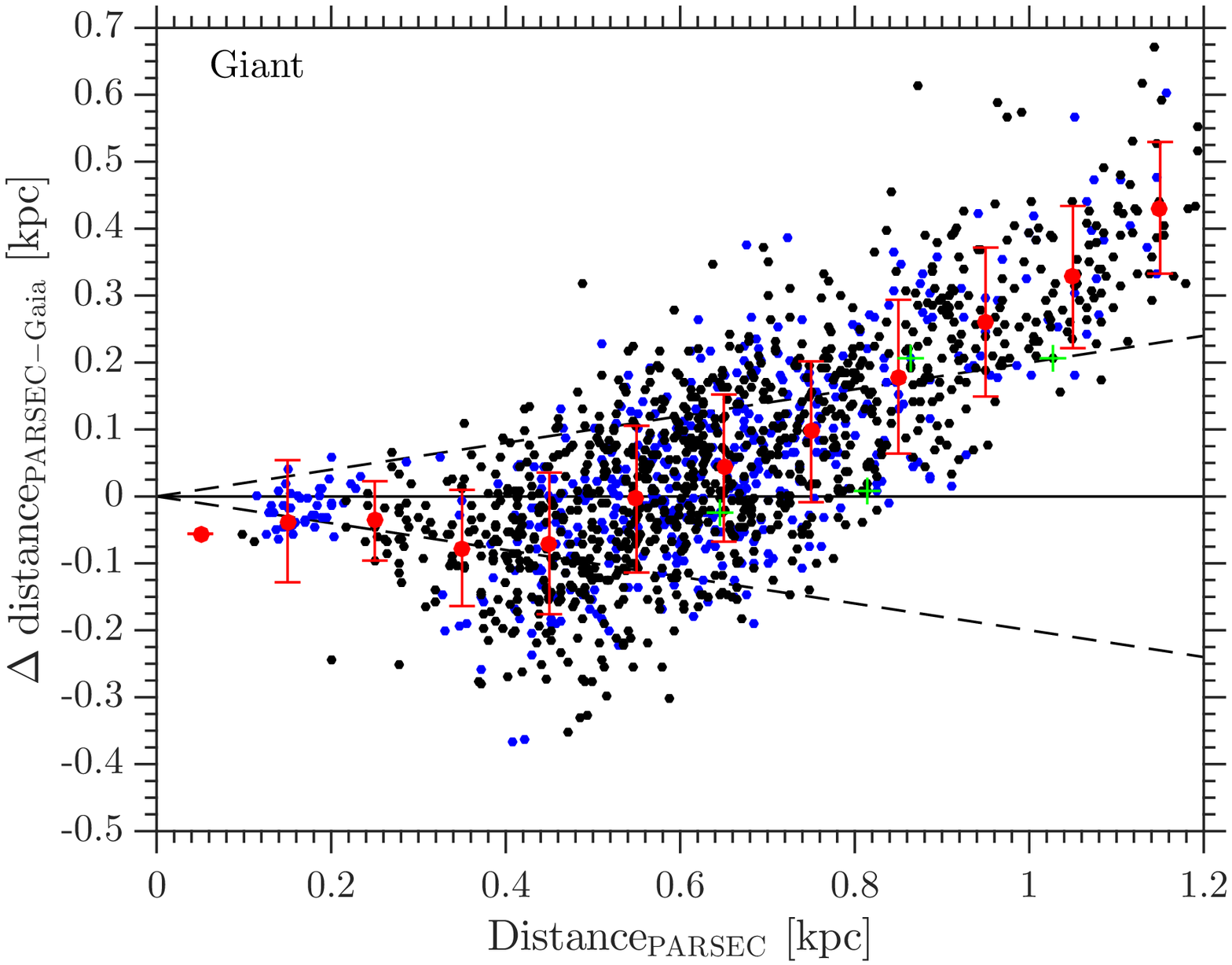}
   \caption{Our model distances compared to those obtained using the TGAS stellar
   parallexes for dwarfs (left) and giants (right). In both panels: the green
   crosses are the sample stars extracted from the $7^\circ \times 7^\circ$ target
   region, and the red dots are the mean values of differences in each distance
   bins with a bin size of 0.1\,kpc. The dash lines indicate the 20\% borders.
   On the right panel: blue dots from APOGEE, and black dots from LEGUE.
   \label{gaiadis}}
\end{figure*}

Figure \ref{gaiadis} compares the distance differences, where the dash lines
delineate the 20\% borders. It can be seen that most dwarfs have the differences
less than 20\%, comparable to the error of TGAS. The mean difference is close to
zero, and a systematical deviation occurs when $d > 0.6$\,kpc in the way the model
distance is larger than the parallactic distance. Dwarfs in our target regions
(green crosses) show a similar tendency. For giants in the right panel of Figure
\ref{gaiadis}, the difference is on the same order as the dwarfs, and has no clear
difference between the LEGUE and APOGEE data. Recalling that the estimated errors
of distances for the LEGUE stars are generally larger than 50\% in Section \ref{stellardis},
the distance errors must be greatly overrated as a result of the overestimation
of $\Delta \logg$ derived from the LAMOST spectra. There is a tendency that the
model distance becomes larger than the parallactic linearly with the distance
when it is greater than 0.6\,kpc. This tendency is visible for both dwarfs and
giants, while more significant for giants at larger distance. That means our
method tends to yield larger distance for relatively distant stars in comparison
with the TGAS data. This may lead to the overestimation of distances. On the
other hand, \citet{dav17} found that the TGAS distance showed systematical deviation 
to larger distance at $d > 0.5$\,kpc for the Kepler field of view. \citet{sta16} 
also reported that the GAIA distance is offset to large. The GAIA distance, when 
$> 0.5$\,kpc, needs better calibration. It's puzzling that the GAIA distance is 
smaller than our model distance when $>0.6$\,kpc. If the problem lies in the model 
distance, the systematic deviation should also occur to the small distance stars 
while it does not.

\section{The Distance and Extinction of the Monoceros SNR} \label{nebdis}

The distance of the Monoceros SNR can now be derived based on the extinction and
distances of individual stars in this sightline. The pre-assumption is that interstellar
extinction increases monotonically with distance at a given sightline, which is
very reasonable as the extinction is an integral parameter along the sightline.
There will be a sharp increase at the position of the Monoceros SNR because of
its higher dust density than the foreground diffuse ISM. The position of the sharp
increase will tell the distance of the nebula.

\subsection{The foreground extinction} \label{foreext}

\begin{figure*}[!htbp]
   \centering
   \includegraphics[scale=0.9]{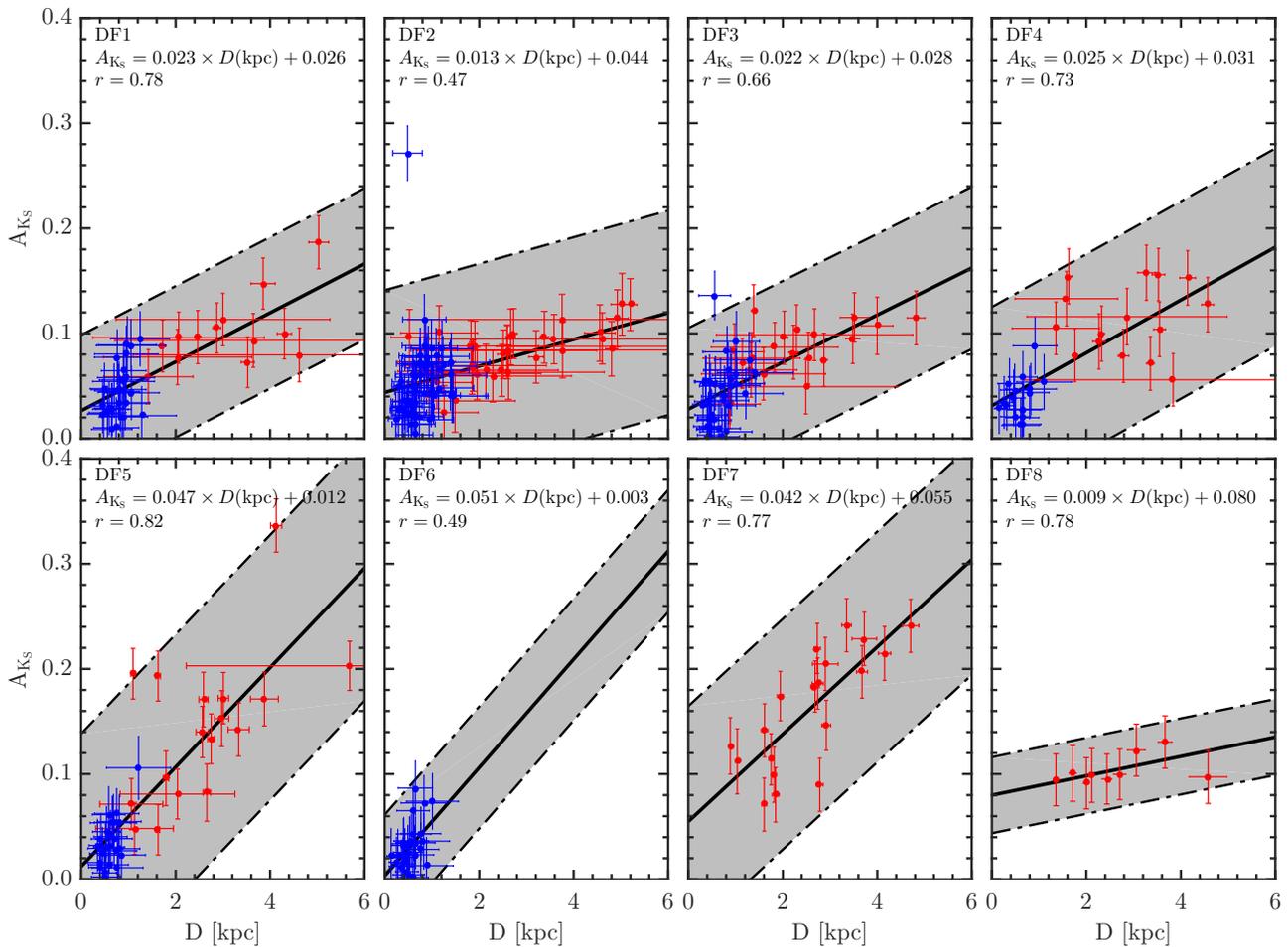}
   \caption{The change of extinction with distance for stars in the 8 DFs. The
   blue dots are dwarfs and the red ones are giants. The fitting results are also
   listed in sub-panels, alongwith the correlation coefficient ($r$) and
   the name of each DF. The grey shaded region in each sub-panels encloses the
   3$\sigma$ uncertainty.
   \label{avdDF8}}
\end{figure*}

Because extinction is an integral effect, the foreground extinction must be
subtracted in order to measure the extinction produced by the SNR alone. For a
precise determination of the foreground extinction, 8 DFs are selected as described
in Section \ref{monarea}. The change of extinction with distance for the stars
in these 8 DFs are shown in Figure \ref{avdDF8} and a linear fitting is performed
for simplicity, alongwith a 3$\sigma$ uncertainty region. It can be seen
that the slopes agree with each other for DF1, DF3, and DF4 with a value of about 
0.02\,mag per kpc in $\AKs$, as well as for DF2 and DF8 with a slightly smaller 
value of about 0.01\,mag per kpc. Meanwhile the DF5 to DF7 variations have a much 
higher slope, being about 0.05\,mag per kpc. This is caused by the Galactic 
latitude as DF1$-$DF4 and DF8 have slightly higher latitude, while the variation 
of slopes between them from 0.009 to 0.025 is mainly due to the local environment.
Considering the average rate of interstellar extinction in the $V-$band is usually 
taken to be $0.7-1.0$\,mag/kpc \citep{got69, mil80}, and the $\Ks-$band extinction 
is about 10\% of the $V-$band, the derived foreground extinction rate does mean a 
diffuse foreground.

\begin{figure}[!htbp]
   \includegraphics[scale=0.9]{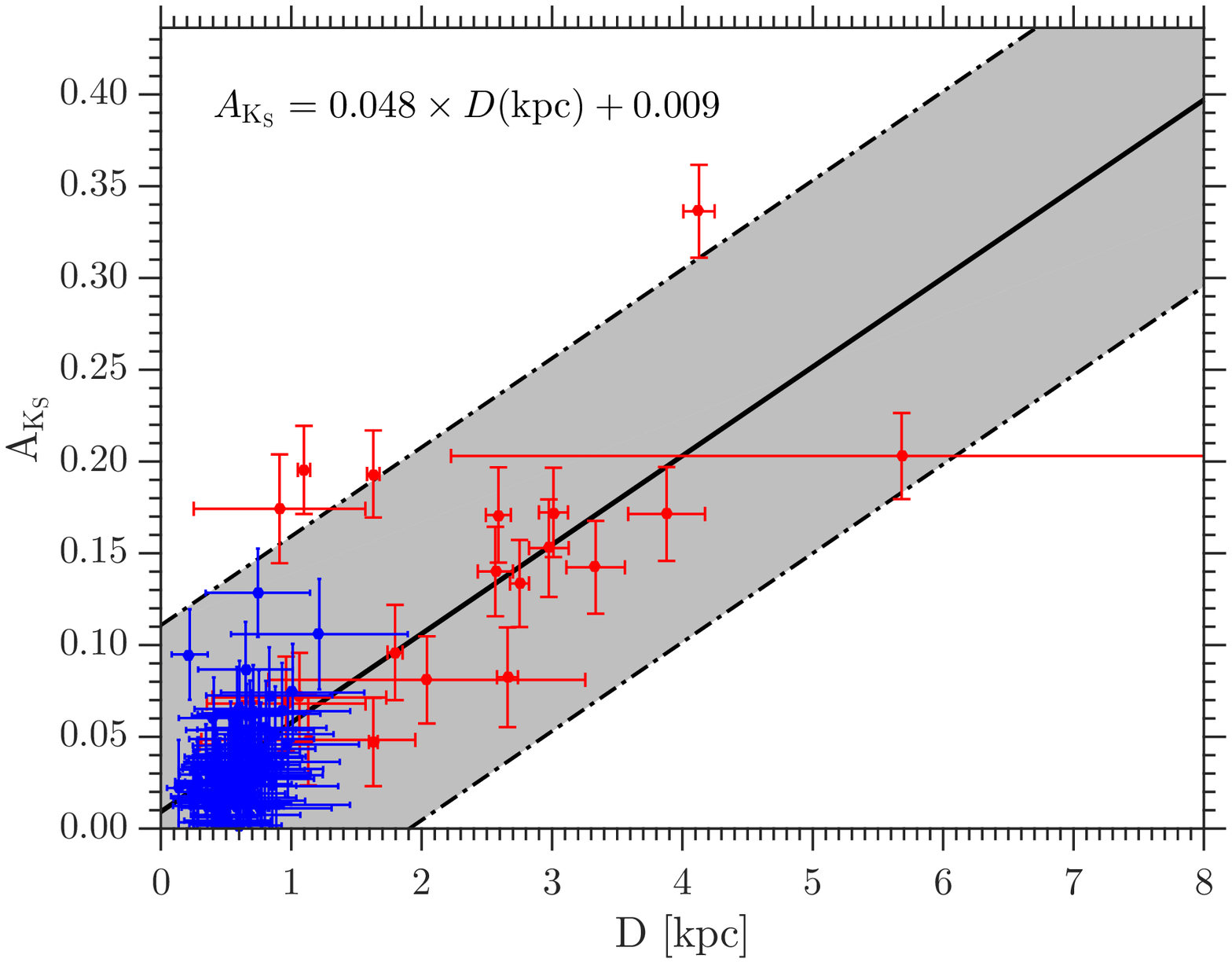}
   \caption{The run of reddening of stars in the reference region, that
   will be used to estimate the extinction contributed by ISD for the Monoceros
   SNR as well as the Rosette Nebula and NGC 2264. The blue dots are dwarfs and
   the red ones are giants. The grey shaded region encloses the 3$\sigma$ uncertainty.
   \label{avdism}}
\end{figure}

For the foreground extinction of the Monoceros SNR, a $1.5^\circ \times 2.0^\circ$
reference region (marked by black dot dash lines in Figure \ref{targets})
is chosen with the center at  $(l,b)=(208\fdg25, +0\fdg5)$ including DF5 and 
DF6, for its similar latitude (Figure \ref{avdism}). This foreground will also be 
applied to the Rosette Nebula and NGC 2264. The extinction of a star, within the 
uncertainty ($3\sigma$) of linear fitting, is mainly produced by the diffuse ISM 
rather than by the nebula. We must take this part of extinction out to study the 
extinction and near-infrared color excess ratios for the nebulas in Section 
\ref{monextlaw}.

\subsection{The Nebular Distance and Extinction} \label{nebuladis}

\begin{figure}[!htbp]
   \includegraphics[scale=0.9]{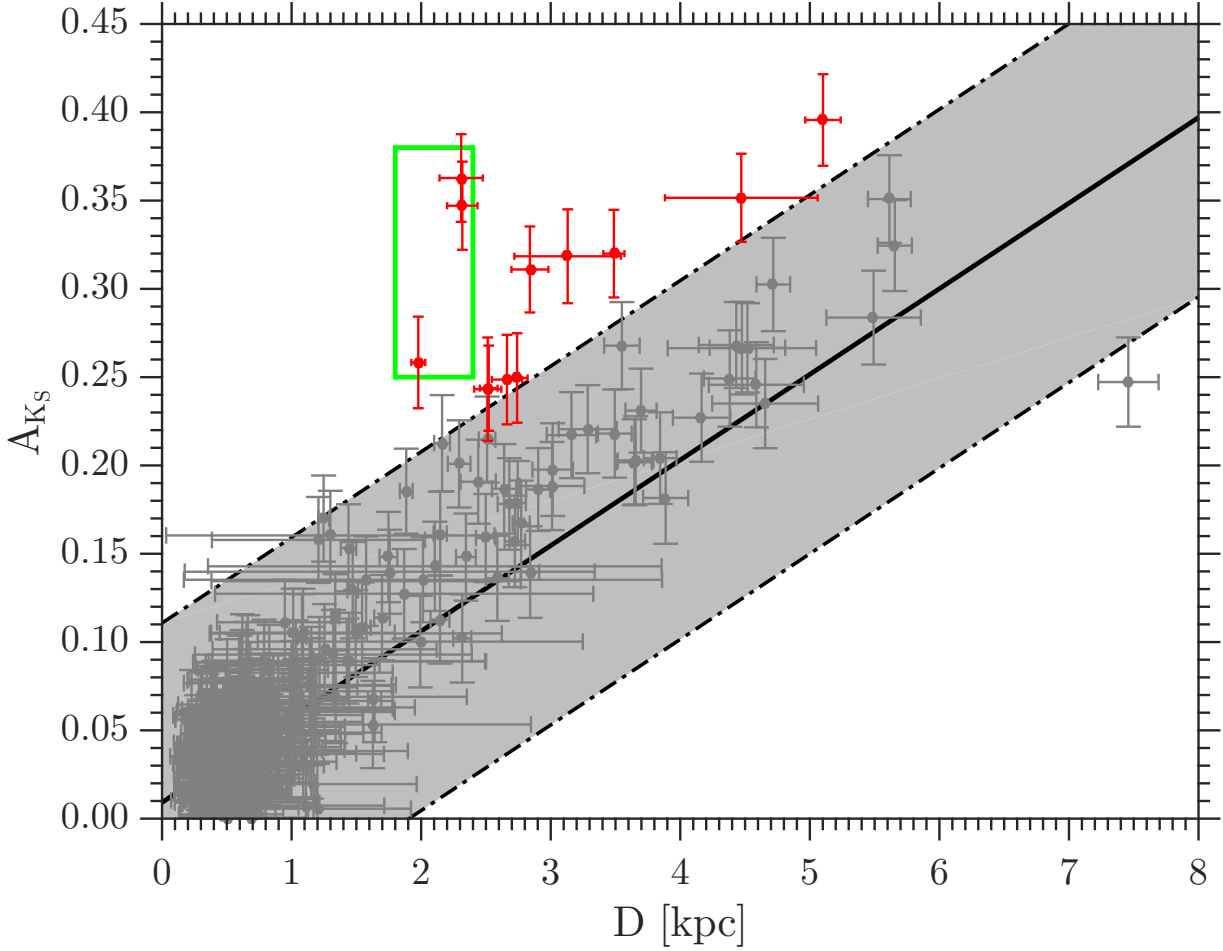}
   \caption{$\AKs$ vs. $D$ for the Monoceros SNR. The background extinction profile
   and the grey shaded region derived from the reference region are the same
   as Figure \ref{avdism}. The red dots are giants which are mainly obscured by
   dust from SNR, while stars in or below the uncertainty region are marked by
   grey dots. The extinction jump can be seen at 1.98\,kpc with $\Delta \AKs \approx
   0.15$ traced by three stars in the green box.
   \label{avdmon}}
\end{figure}

\begin{figure}[!htbp]
   \includegraphics[scale=0.9]{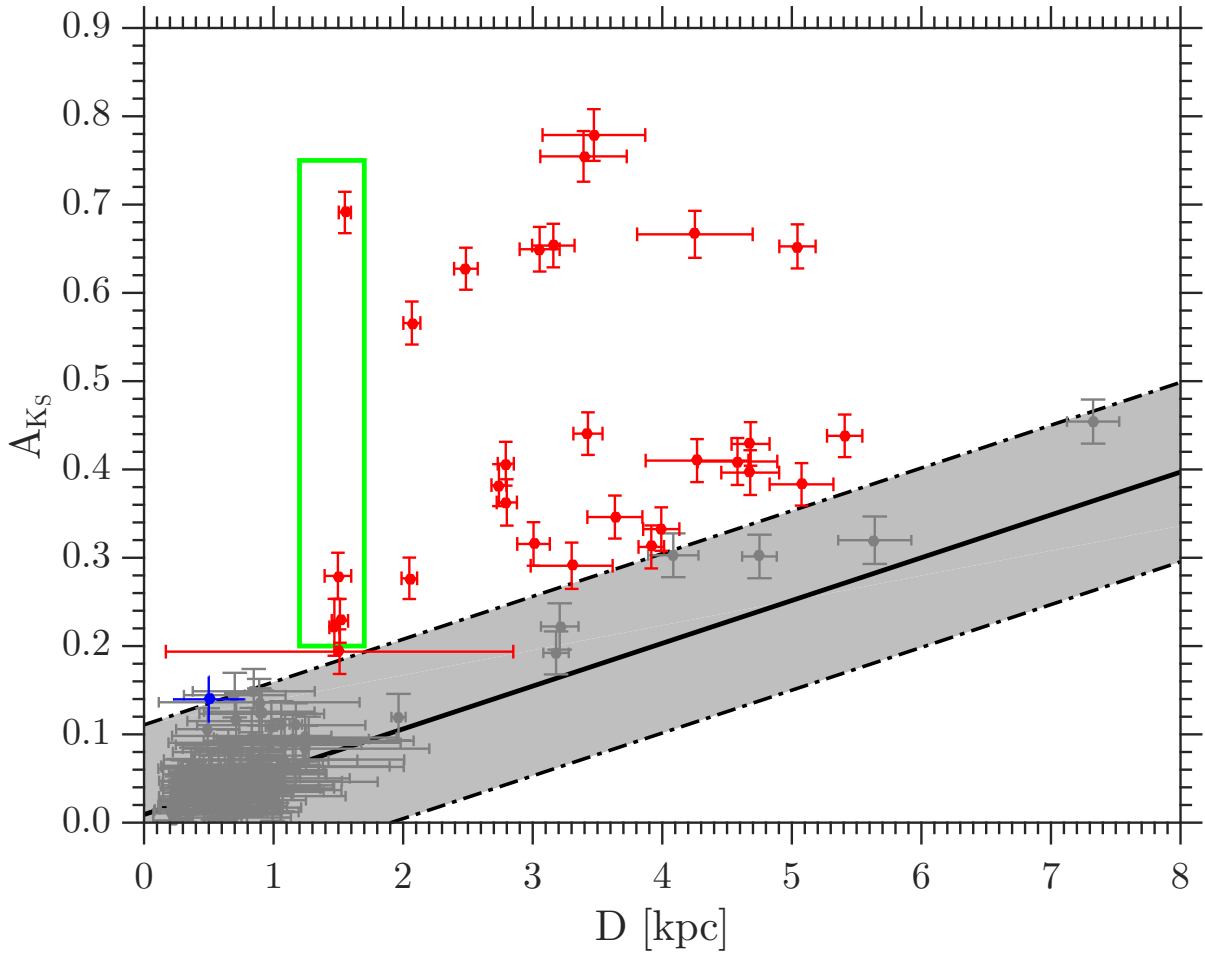}
   \caption{The same as Figure \ref{avdmon}, but for the Rosette Nebula. The jump
   of $\AKs$ can be clearly seen at 1.55\,kpc, followed by several high-extinction
   stars in the green box. The blue point represents a dwarf above the uncertainty
   region.
   \label{avdros}}
\end{figure}

\begin{figure}[!htbp]
   \includegraphics[scale=0.9]{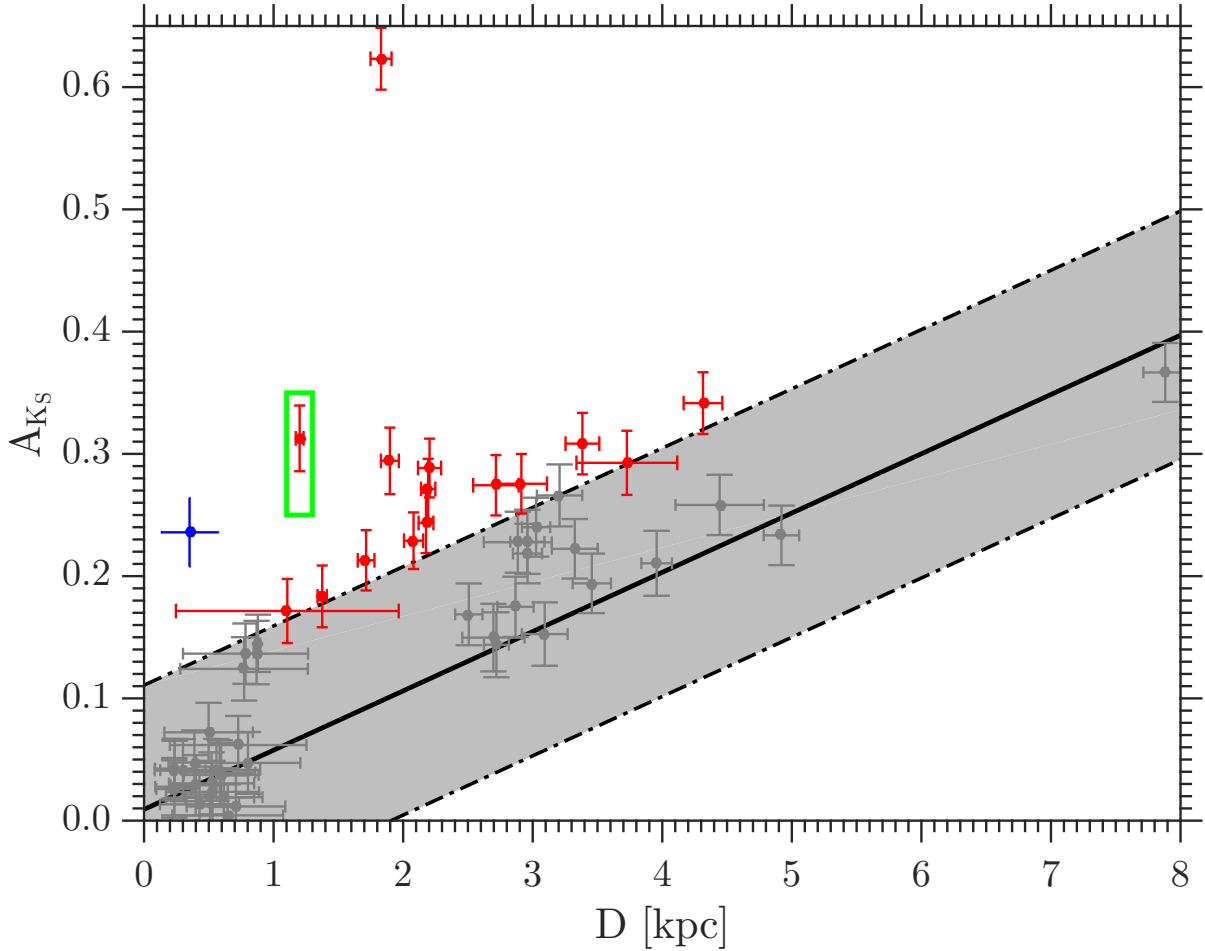}
   \caption{The same as Figure \ref{avdmon}, but for NGC 2264.
   \label{avdngc}}
\end{figure}

The change of stellar extinction $\AKs$ with distance $D$, is shown in Figures
\ref{avdmon}$-$\ref{avdngc} for the three selected nebular regions. To be reasonable,
only stars with $\EJH>0$ and $\EJKs>0$ are regarded as the correct indicators.
For a better accuracy, $\sigma_{\rm D} / D <100\%$ is also required. 
For the Monoceros SNR, it can be seen that there are three stars (located in the 
green box in Figure \ref{avdmon}) whose extinctions clearly jump around 2.0\,kpc. 
In order of distance, they are (1) $\AKs = 0.26$ at 1.98\,kpc, (2) $\AKs = 0.36$ 
at 2.31\,kpc, and (3) $\AKs = 0.35$ at 2.32\,kpc. As the nebular extinction shows 
up only when the star lies behind, the stellar distance should be the upper limit 
of the Monoceros SNR. The three stars thus indicate the upper limit of the distance. 
We tend to believe the closest distance, i.e. 1.98\,kpc is the nebular distance 
and the other two stars are behind the SNR. The dispersion of the extinction is 
mainly caused by the inhomogeneity of the SNR. On the other hand, the tracers are 
located densely around 2.0\,kpc, this distance should be very close to the position 
of the SNR nebula. In addition, there is no apparent increase of extinction up 
to at least 1.9\,kpc. Therefore, the distance of the Monoceros Nebula is between 
$1.90-1.98$\,kpc.

The extinction of the Rosette Nebula, $\Delta \AKs \approx 0.5$\,mag, is twice
that of the Monoceros SNR. From Figure \ref{avdros}, the distance of Rosette can
be determined to be less than 1.55\,kpc as a star at 1.55\,kpc has an apparent
increase in $\AKs$, with $\Delta \AKs > 0.5$\,mag, which is followed by several
stars (in the green box in Figure \ref{avdros}) with similarly steeply rising
extinction. NGC 2264 has an extinction jump of $\Delta \AKs \approx 0.25$\,mag
at $1.20 \pm 0.03$\,kpc (Figure \ref{avdngc}), which sets the distance at 1.20\,kpc.
However, there is one dwarf (blue cross in Figure \ref{avdngc}) with a distance
of 0.35\,kpc and $\AKs = 0.24$\,mag, obviously larger than other dwarfs nearby.
We suspect this star is mis-classified as a dwarf while it may be a giant star
at much larger distance. No cloud is claimed at this distance at this sightline.
In addition, no neighbour stars follow the tendency, and this distance is too much
smaller than previous results. Instead, there are quite some stars showing up above
the foreground and background extinction after the star at 1.20\,kpc. So 1.20\,kpc
should be the distance of NGC 2264.

Table \ref{nebdistab} compares the derived distances to the three nebulas with
previous studies. The distance of the Monoceros SNR is 1.98\,kpc, appearing larger
than previous value of $\sim$1.6\,kpc. Meanwhile, the distance of the Rosette
Nebula, 1.55\,kpc, coincides with previous results. According to our new determinations
of the distances, there should be no interaction between these two nebulas as
their distance difference is about 0.4\,kpc. The distance to NGC 2264, 1.2\,kpc,
is larger than previous results, but quite close to the result of \citet{mor65},
0.95\,kpc. Overall, the positional relation of the three nebulas is consistent
with \citet{dav78}, i.e. the Monoceros Nebula is the furthest, NGC 2264 the closest
and the Rosette Nebula in-between.

\begin{figure*}[!htbp]
   \centering
   \includegraphics[scale=0.30]{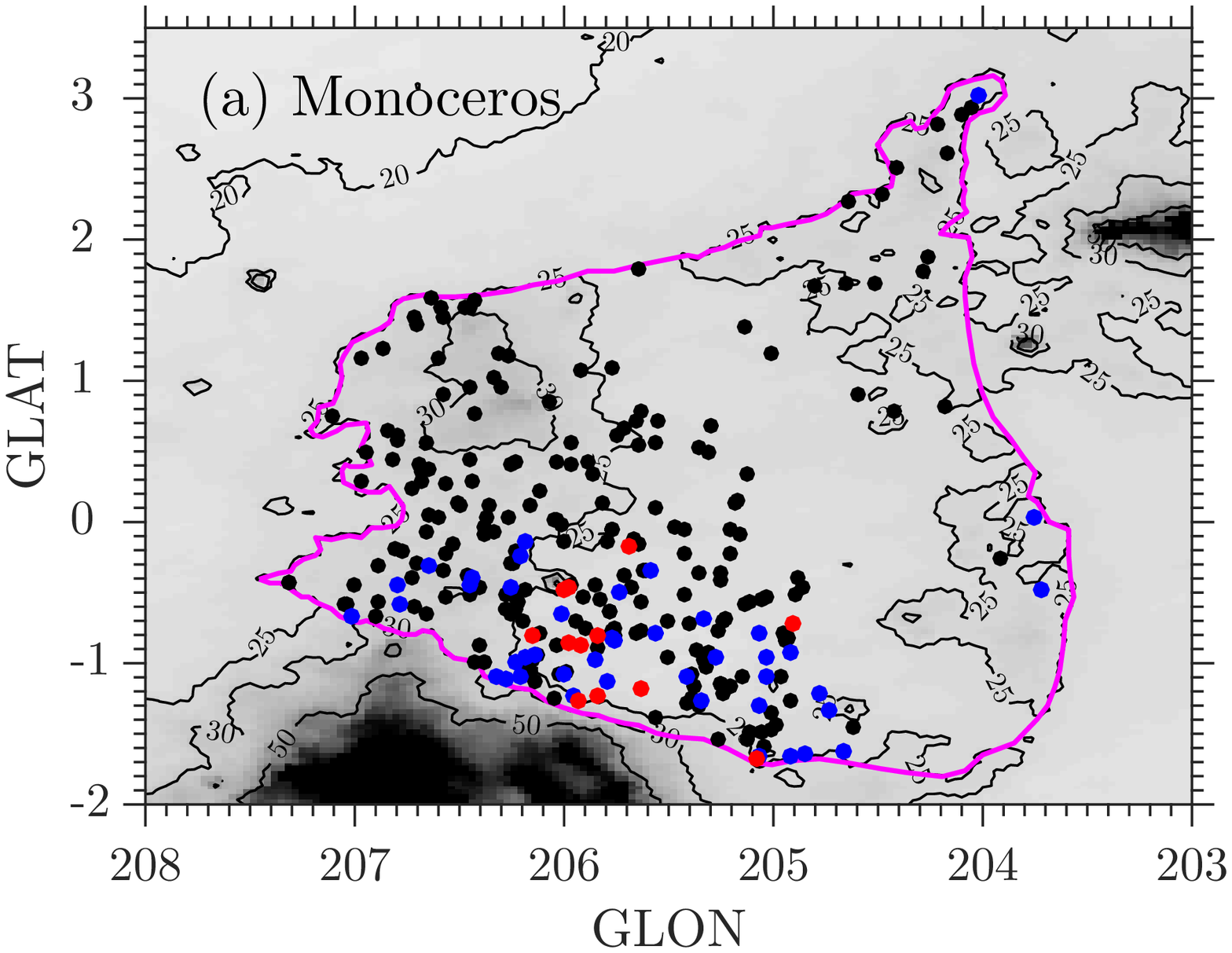}
   \includegraphics[scale=0.30]{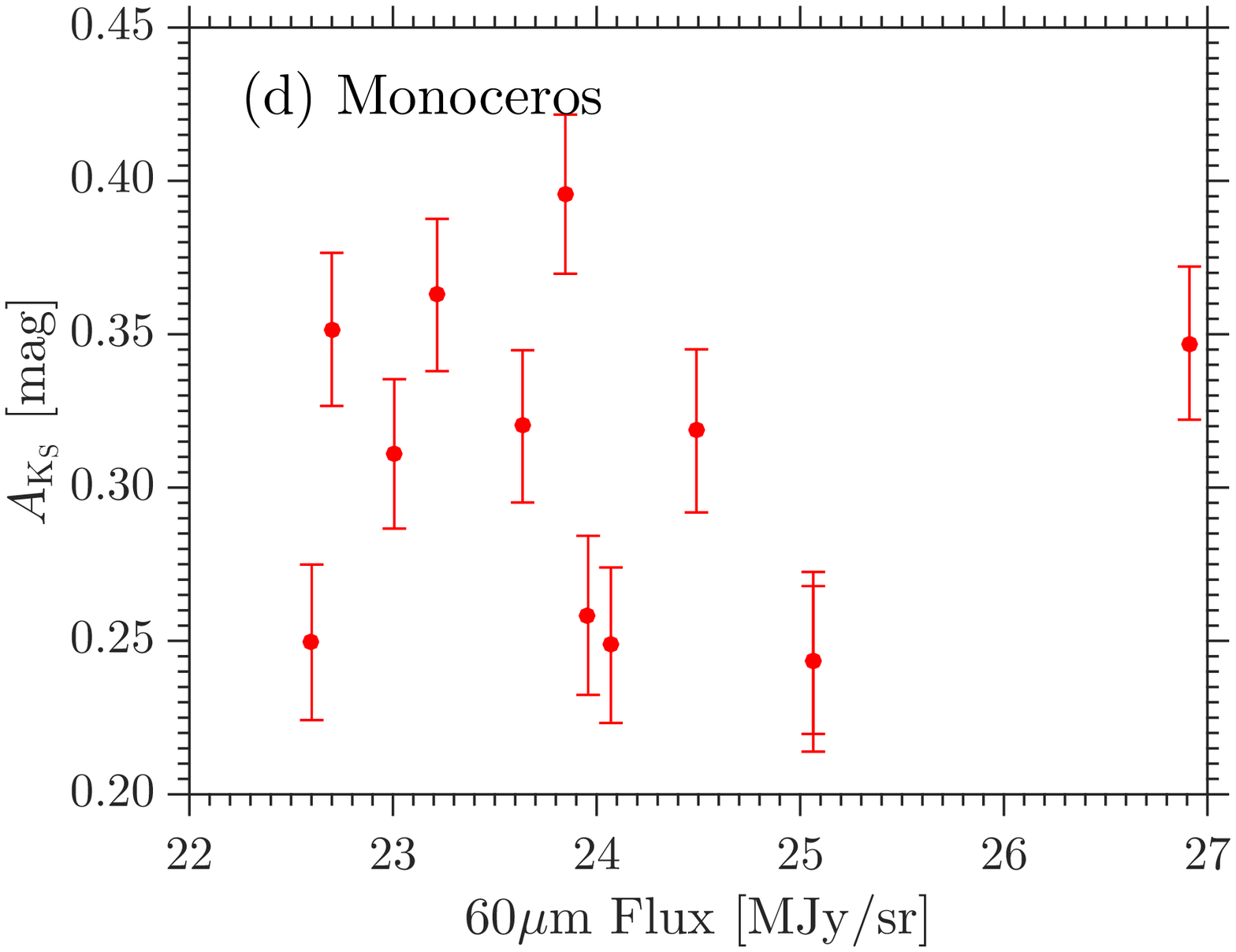}
   \includegraphics[scale=0.30]{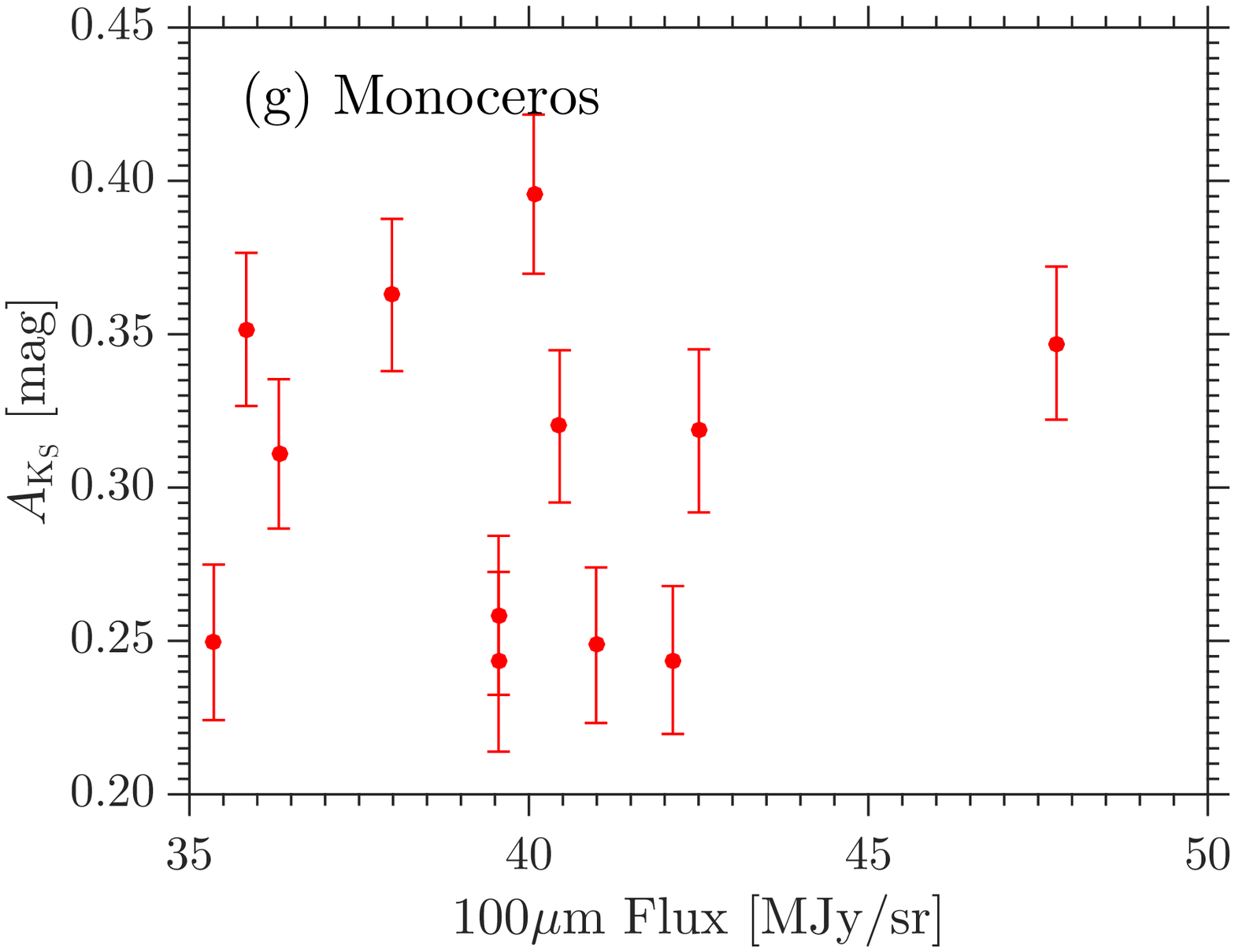}
   \includegraphics[scale=0.30]{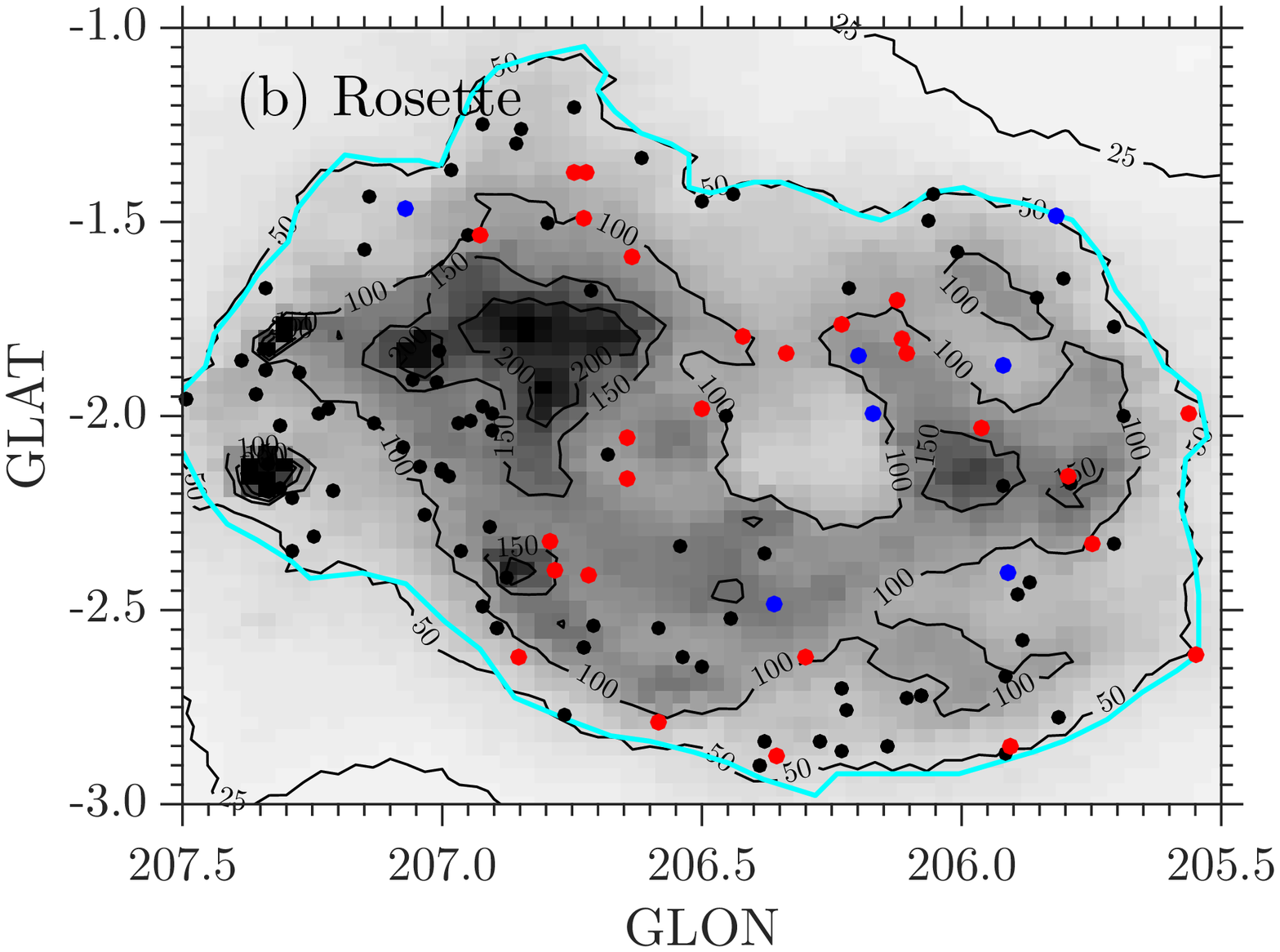}
   \includegraphics[scale=0.30]{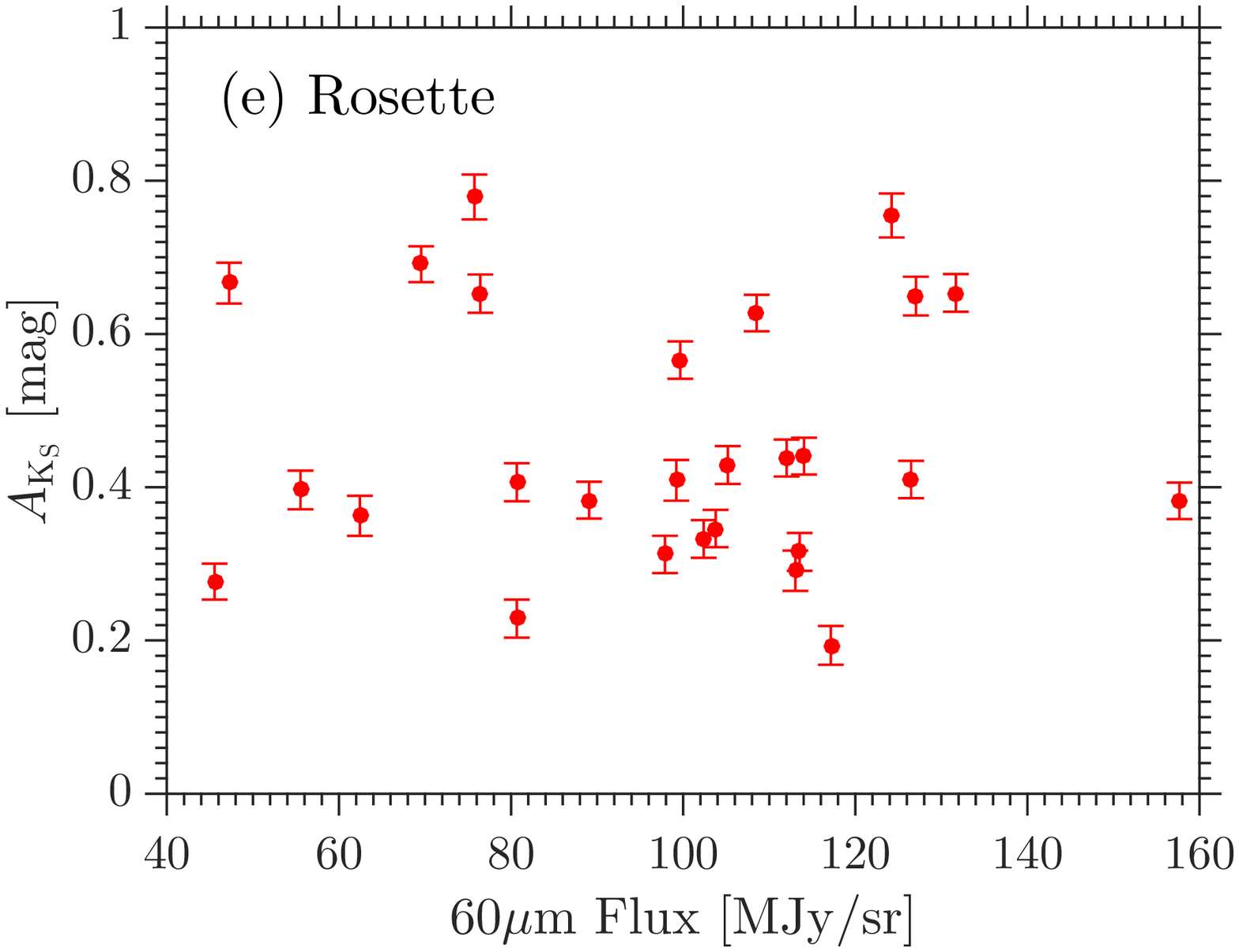}
   \includegraphics[scale=0.30]{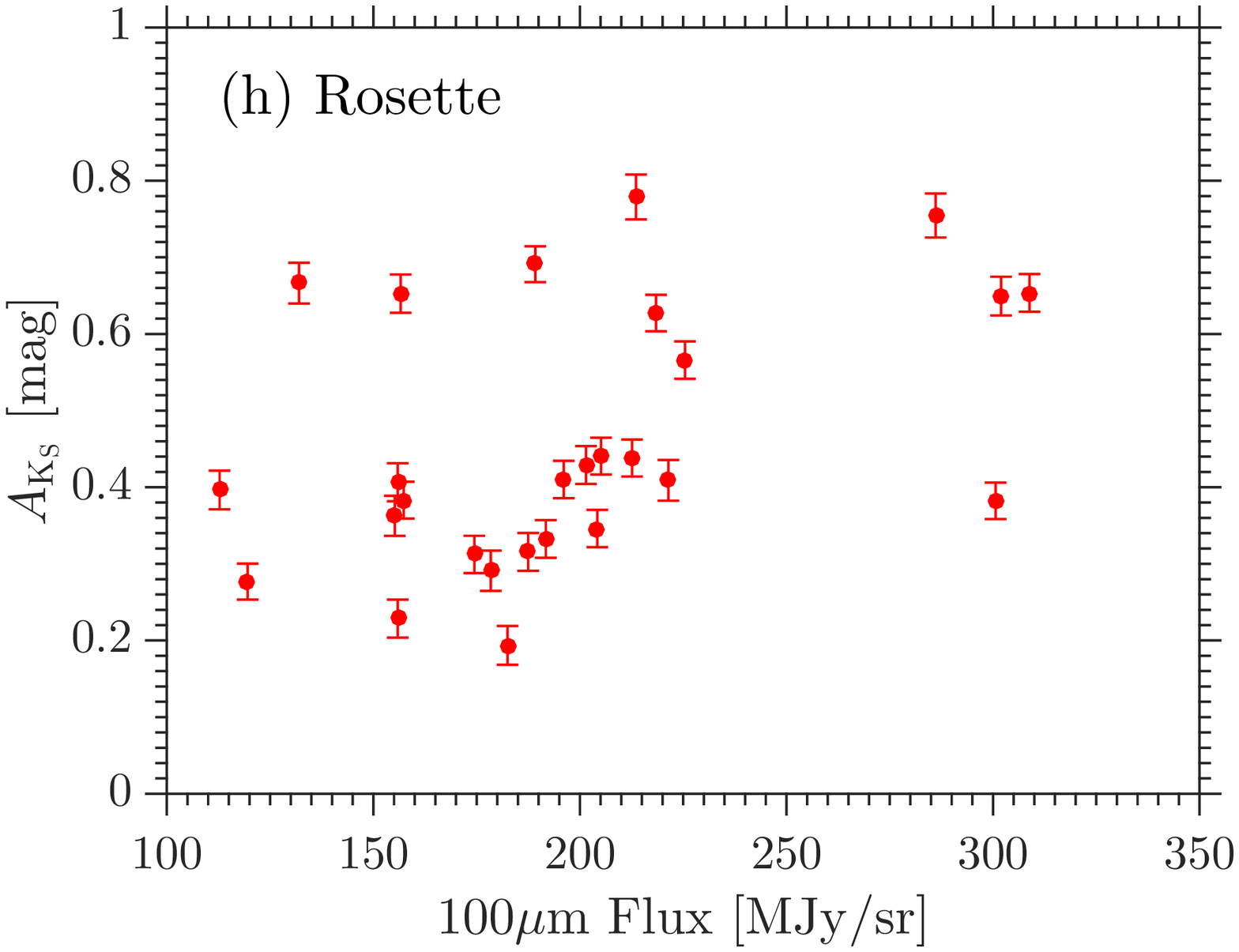}
   \includegraphics[scale=0.30]{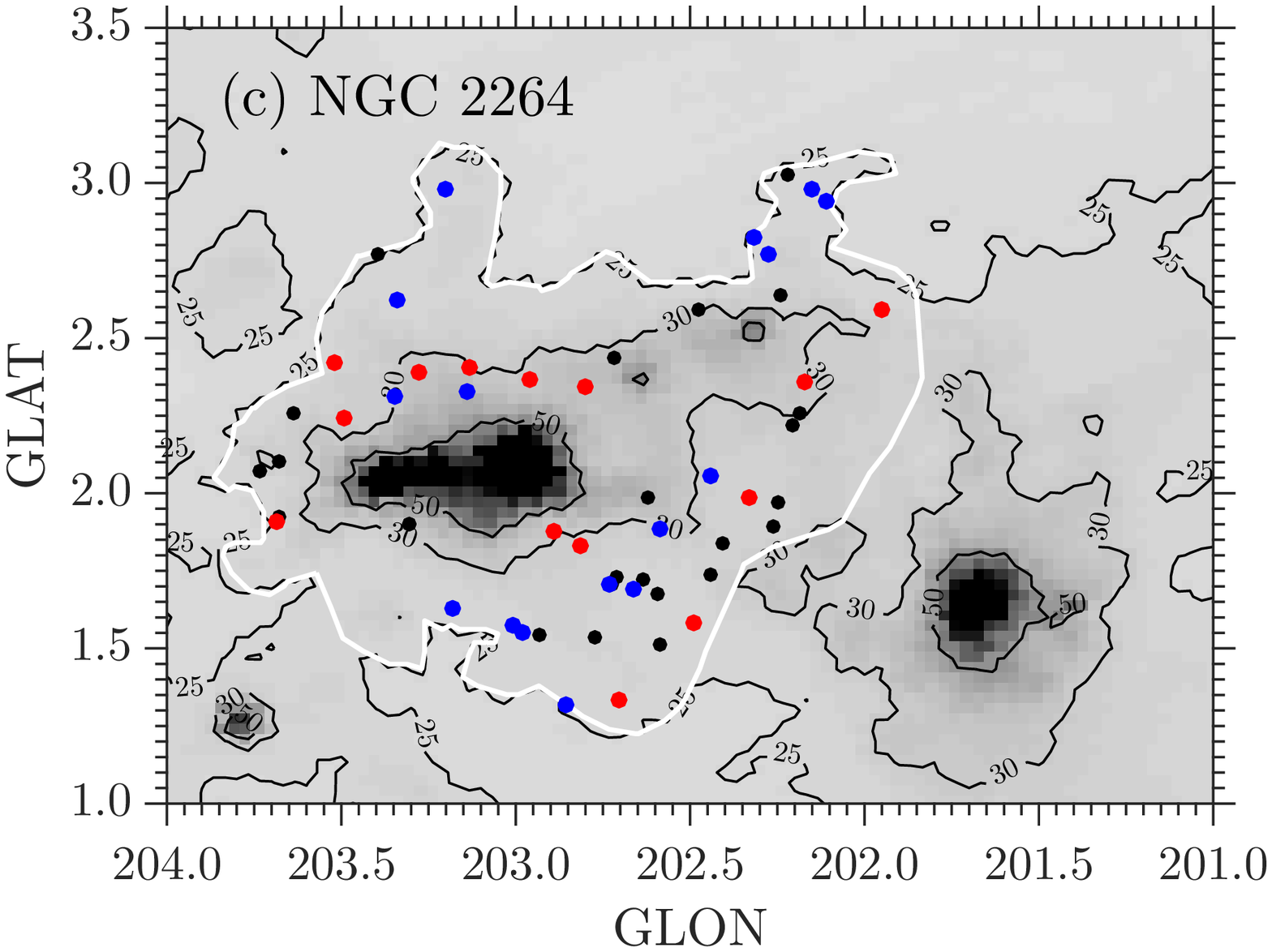}
   \includegraphics[scale=0.30]{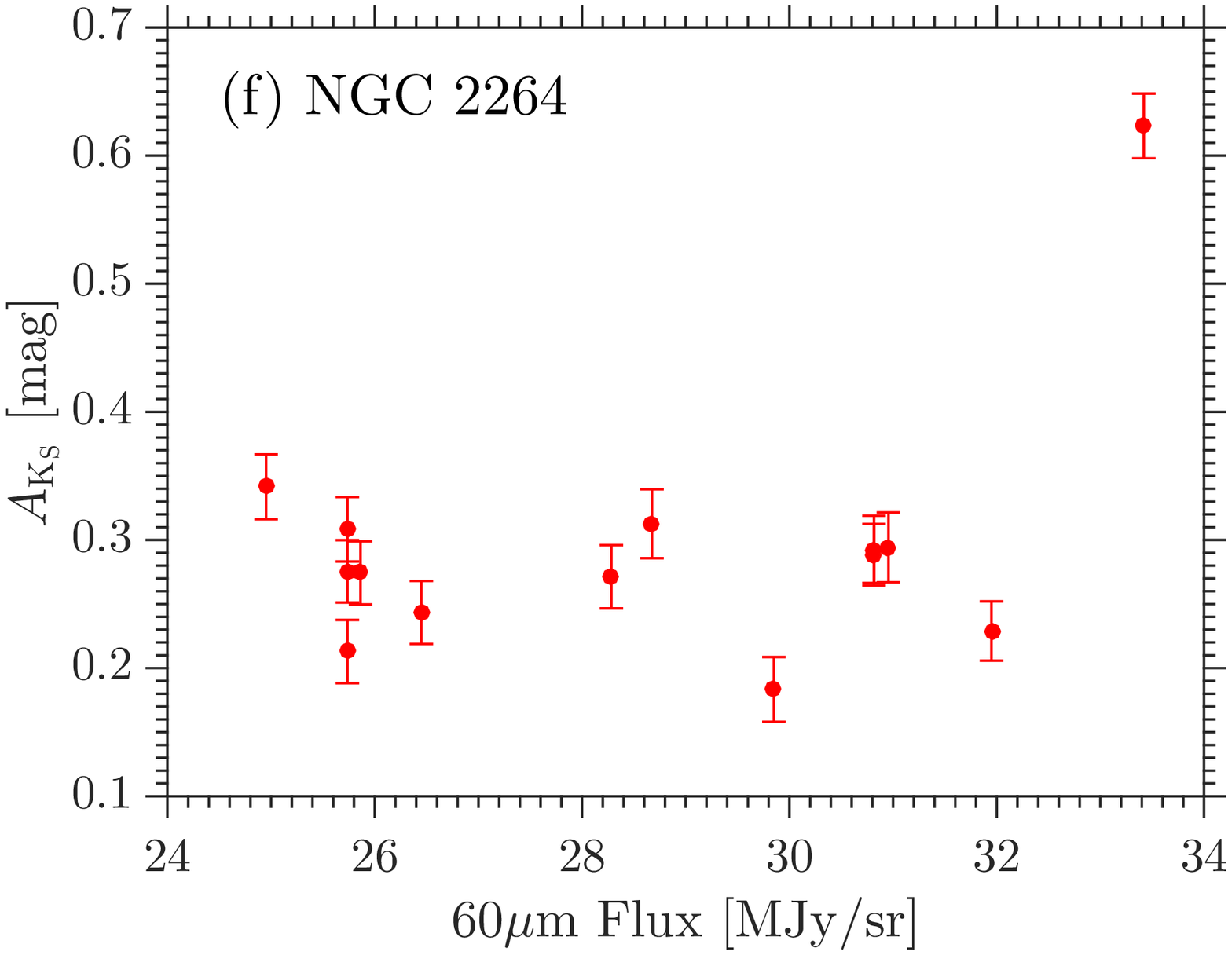}
   \includegraphics[scale=0.30]{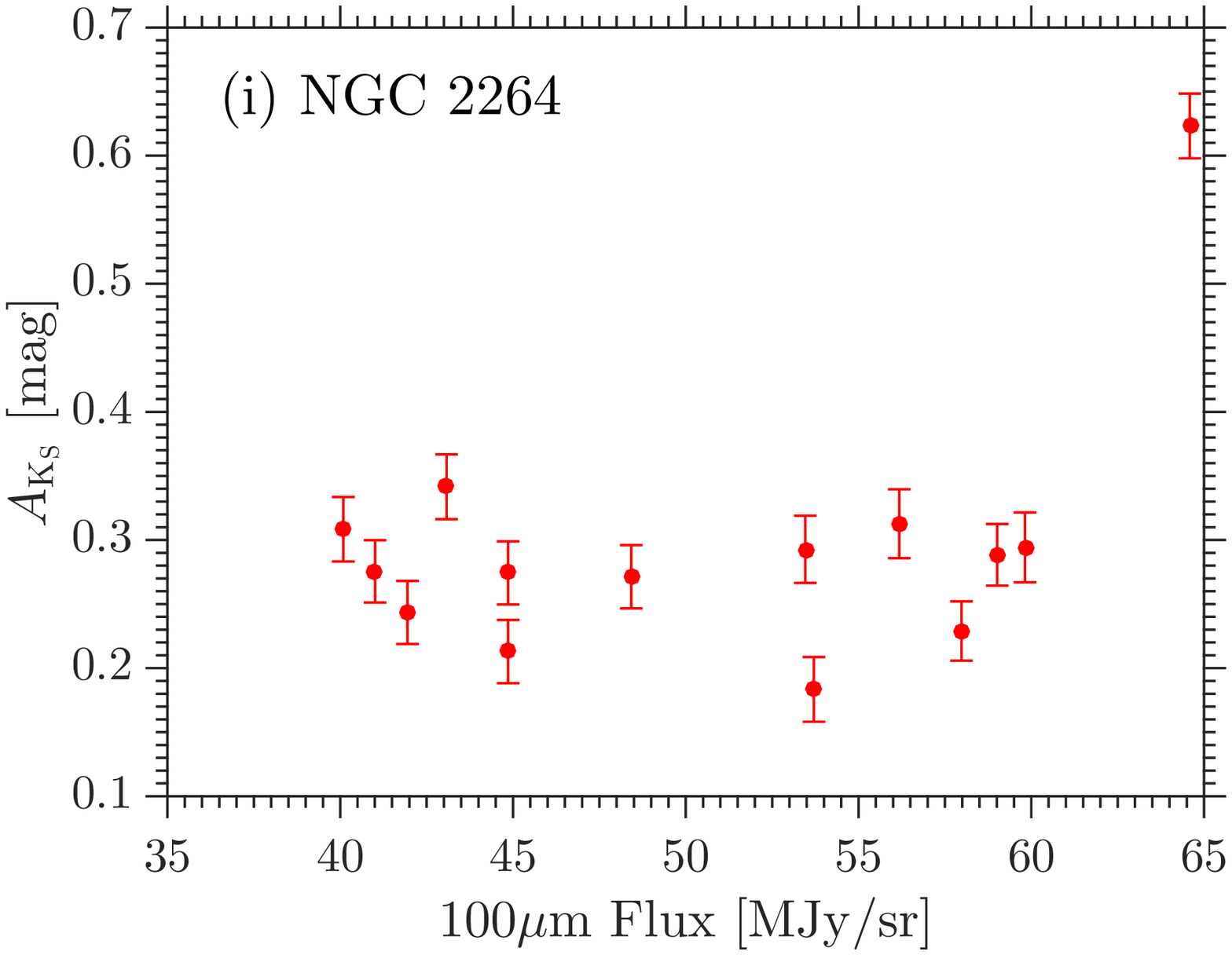}
   \caption{Left panels: The distribution of sample stars. The red dots
   are the tracing stars, the black ones are the foreground stars, and the blue
   ones are the stars behind the nebulae which are mainly obscured by the interstellar
   dust. The contours and nebular borders are the same as Figure \ref{targets}. Middle
   and right panels: The relationship between the extinction, $\AKs$, and the 60 and
   100\,$\mu m$ flux from IRAS for the three nebulae, respectively.}
   \label{triDis}
\end{figure*}

The location of the nebular tracers is shown in Figure \ref{triDis}\,(a)$-$(c).
There are no stars in the highest 60$\mu m$ emission regions for all the three 
nebulas very possibly because of too high extinction in comparison with the depth 
of observation. The tracers mainly distribute near the southern edge of the Monoceros 
SNR, while the foreground stars with low extinction spread in a wide distance 
range. No extinction jump is found for these foreground stars in Figure \ref{avdmon}, 
which indicates that the sharp increase in the extinction at $1.90-1.98$\,kpc 
can only be attributed to the SNR. Although tracers of the Rosette Nebula are 
more scattering, the crucial ones still have nearby foreground stars to ensure 
the distance estimation. As for NGC 2264, with fewer stars, it is hard to exclude 
the existence of a foreground cloud. But NGC 2264 itself contains a massive dark 
cloud and the previous work implies a nearest distance of 0.8\,kpc, so the possibility 
is low for a comparable dust cloud in a nearby region.

\begin{figure*}[!htbp]
   \centering
   \includegraphics[scale=0.45]{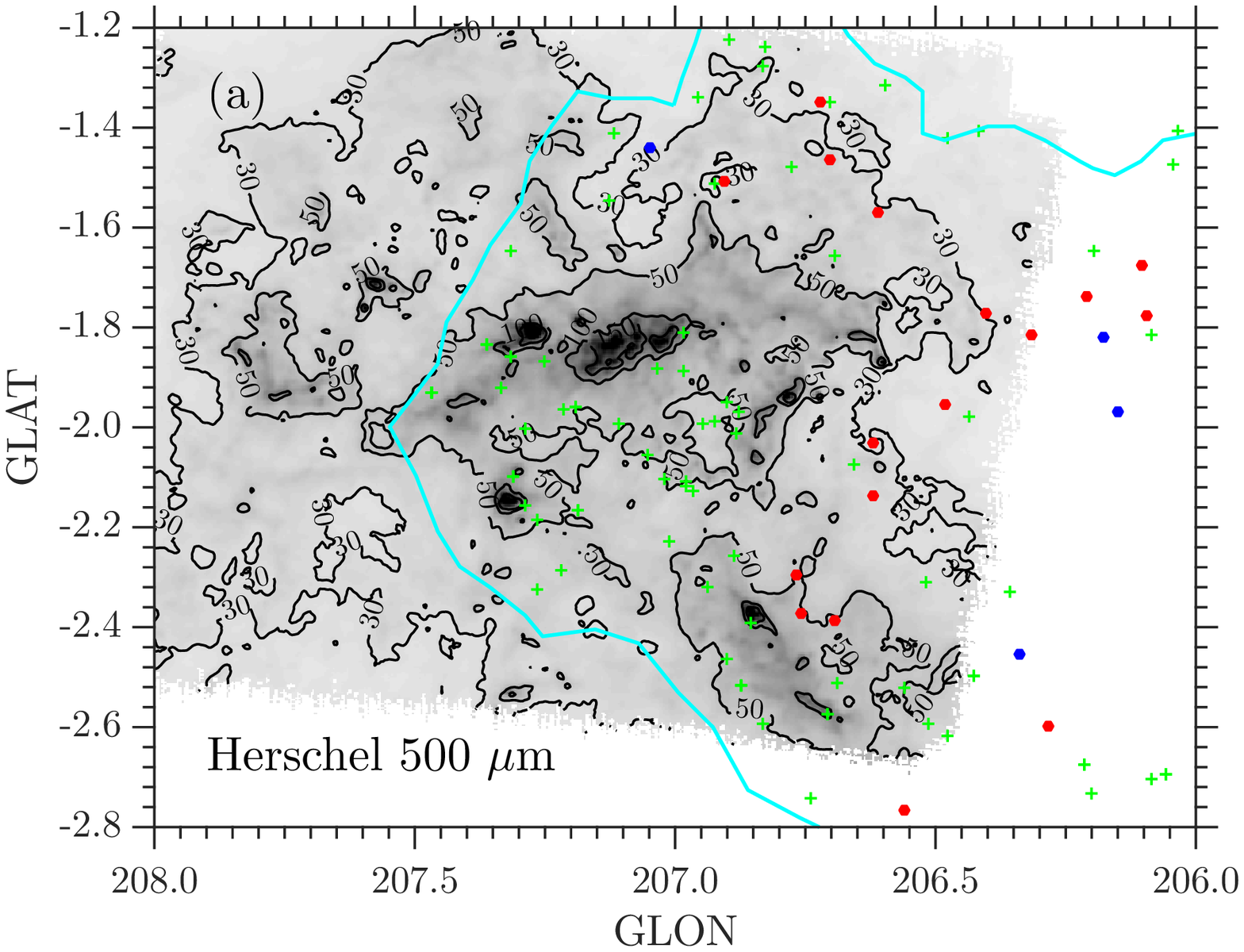}
   \includegraphics[scale=0.45]{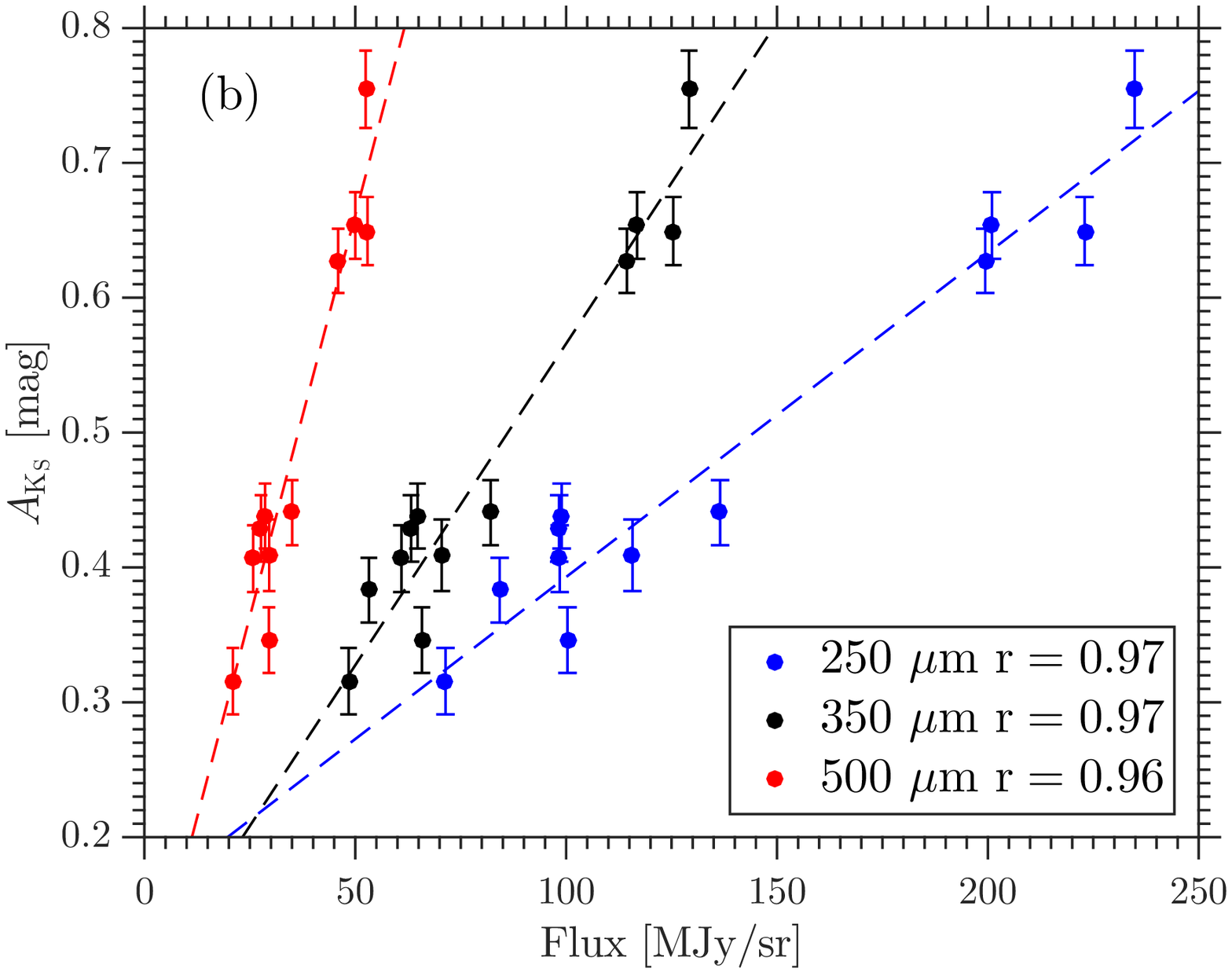}
   \caption{Left: The SPIRE 500\,$\micron$ image of part of the Rosette
   Nebula with our sample stars in this sightline and the nebular border (the 
   cyan profile). The green crosses are the foreground stars, and the red and 
   black dots are the same as in Figure \ref{triDis}. Right: The linear relationships
   between the extinction $\AKs$ and the infrared emission intensity of the Rosette
   Nebula at 250, 350, and 500\,$\micron$, respectively. The correlation coefficients 
   ($r$) are shown in the legend box.}
   \label{HRose}
\end{figure*}

The nebular dust not only causes extinction to the background stars, but 
also emits infrared radiation, thus a correlation between the nebular extinction 
and infrared emission is expected. Figure \ref{triDis} compares the extinction of 
stars behind the nebulas and the infrared flux of the nearest pixel as per the IRAS
60\,$\micron$ (middle panels) and 100\,$\micron$ (right panels) image, respectively.
We made no intention to subtract the background emission from the infrared images
because it would be non-uniform for a large extended nebula, such as Monoceros,
and consequently hard to model. No correlation is found between the extinction 
and the 60\,$\micron$ emission or the 100\,$\micron$ emission. Although both the 
extinction and emission is proportional to dust mass, the emission depends 
sensitively on dust temperature. The 60\,$\micron$ and 100\,$\micron$ emission 
is dominated by warm dust that makes up only a small fraction of the total dust 
in SNRs (see the dust mass estimation of \citet{gom12} and \citet{loo17}). It 
also implies that the warm and cold dust do not spatially coincide completely, 
which is suggested by the dust map of \citet{loo17}. A check of the dust emission 
at longer wavelength may reveal whether the excess extinction is due to the 
nebular dust. Fortunately, the eastern part of the Rosette Nebula was observed 
by the Herschel Space Observatory \citep[HSO;][]{pil10}, with its Spectral and 
Photometric Imaging Receiver \citep[SPIRE;][]{gri10} at 250, 350, and 500\,$\micron$. 
We obtained the reduced SPIRE data through the Herschel Interactive Processing 
Environment \citep[HIPE;][]{ott10}. Figure \ref{HRose}\,(a) shows the Herschel 
500\,$\micron$ image of the Rosette Nebula together with the sample stars and 
the nebula border. Most tracing stars are located in the region with the intensity 
of $30-50$\,MJy/sr, while the dense region is not covered again due to its severe 
extinction. The distances of individual stars and background emission have much 
smaller influence at far-infrared that is dominated by the nebular cold dust. It 
can be seen that there exists tight linear relations of nebular stellar extinction, 
$\AKs$, with the dust emission at 250, 350, and 500\,$\micron$, respectively (Figure 
\ref{HRose}\,(b)), which yields the linear correlation coefficient greater than 0.96. 
This result shows that the extinction-producing dust is identical to the far-infrared 
emission dust.

\begin{table*}
\begin{center}
\caption{\label{nebdistab} The nebular distances (in kpc) compared with previous
works.}
\begin{tabular}{lccc}
\tableline \tableline
                & Monoceros SNR & Rosette Nebula & NGC 2264 \\
\tableline
  This work (upper limit) &  1.98 & 1.55 &  1.2 \\
  \citet{joh62} & -   & 1.66 & -     \\
  \citet{BF63}  & -   & 2.2  & 0.715 \\
  \citet{mor65} & -   & 1.7  & 0.95  \\
  \citet{dav78} & 1.6 $\pm$ 0.3 & 1.6 & 0.8 \\
  \citet{gra82} & 1.6 & -    & -     \\
  \citet{lea86} & 1.5 & -    & -     \\  
  
\tableline
\end{tabular}
\end{center}
\end{table*}

\section{The Near-Infrared Extinction Law} \label{monextlaw}

Although the near-infrared extinction law takes the form of a power law, the power
index $\alpha$ is very sensitive to the adopted wavelengths of the $JH\Ks-$bands.
So the color excess ratio, $\EJHJKs$, is a more stable and reliable description
of the near-infrared extinction law. \citet{WJ14} and \citet{xue16} have already
derived the mean $\EJHJKs$ of the Milky Way, which are 0.64 and 0.652 respectively
and consistent with each other, and the result by \citet{xue16} is more preferable
for their better determination of the intrinsic color indexes.

\begin{figure*}[!htbp]
   \centering
   \includegraphics[scale=0.45]{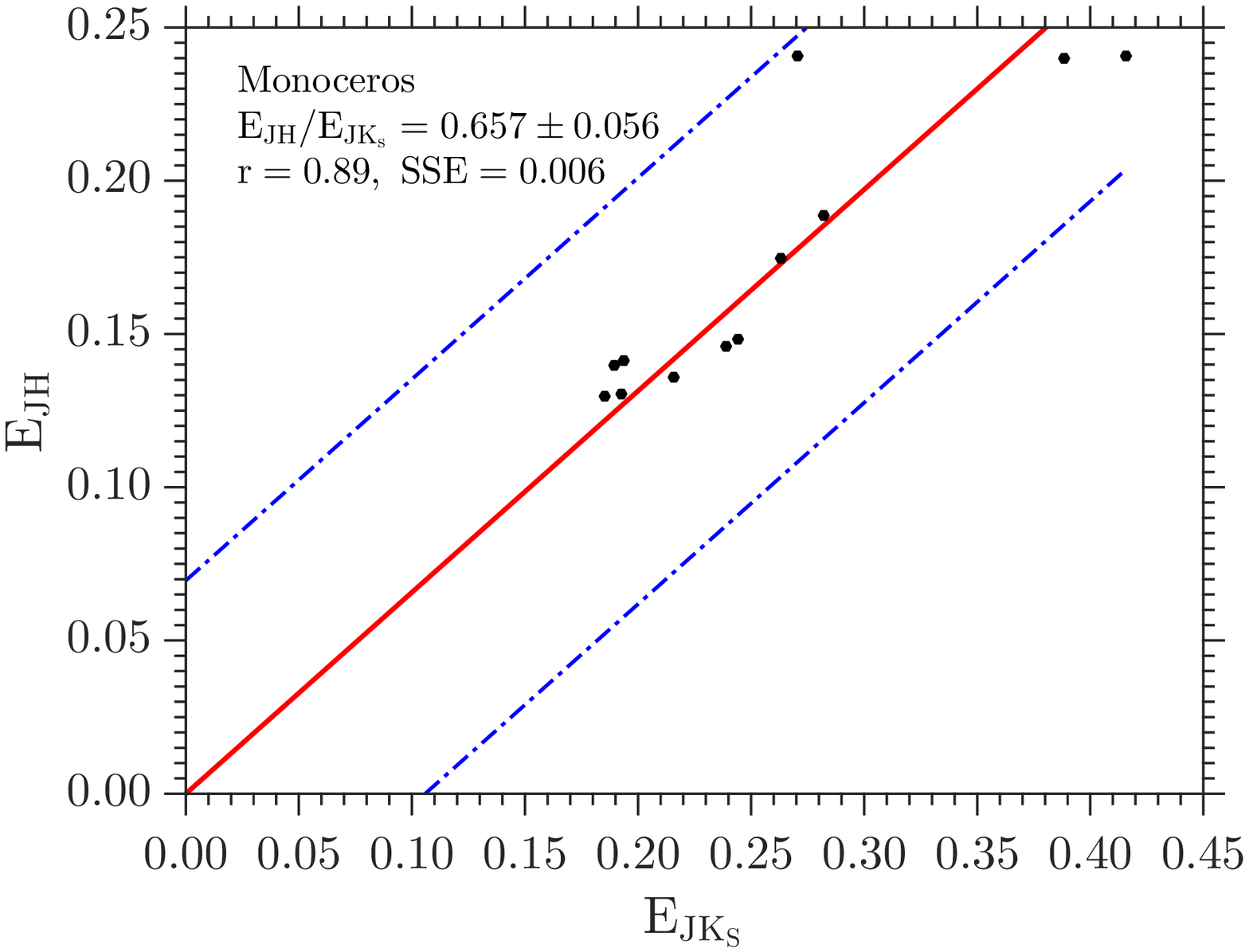}
   \includegraphics[scale=0.45]{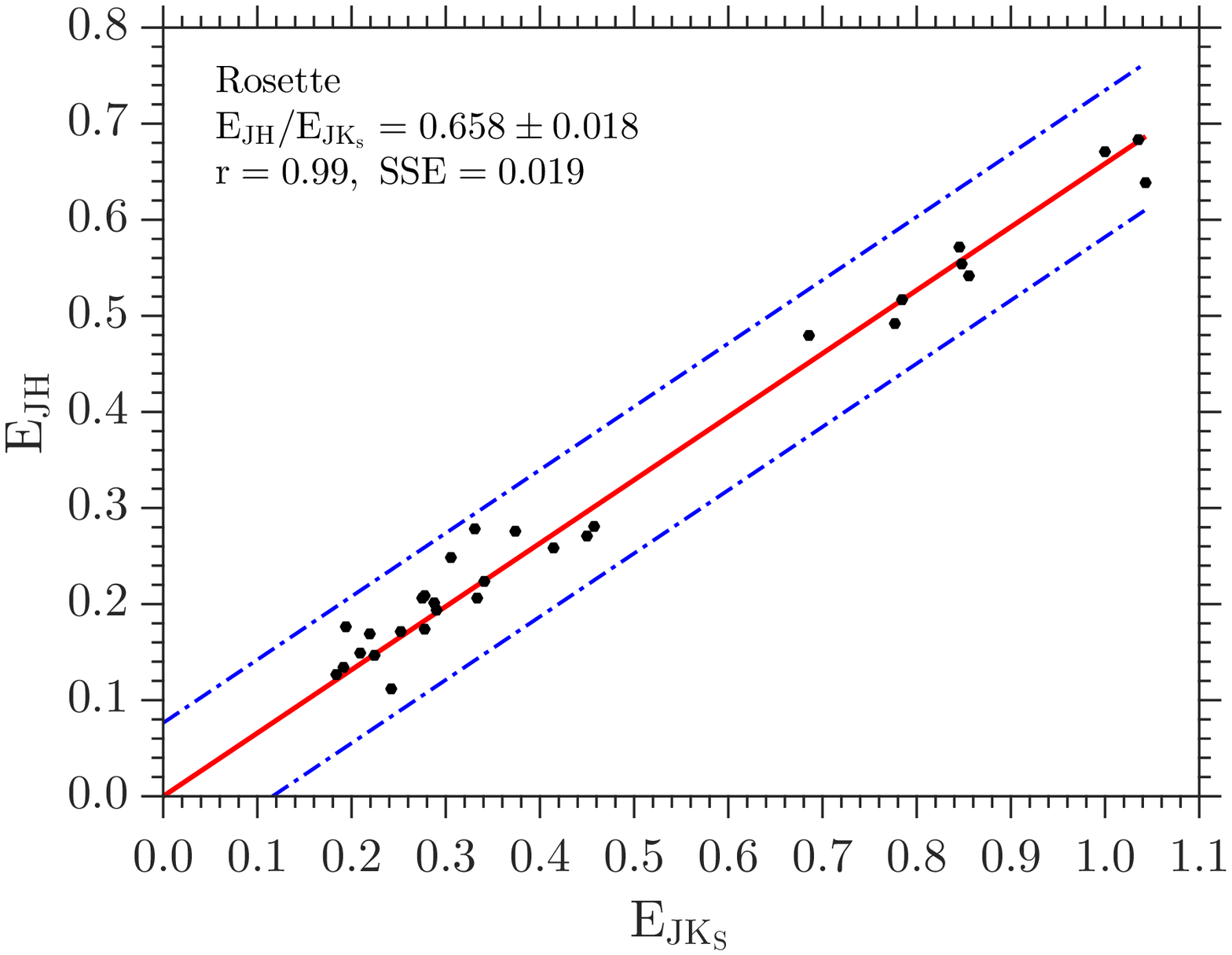}
   \includegraphics[scale=0.45]{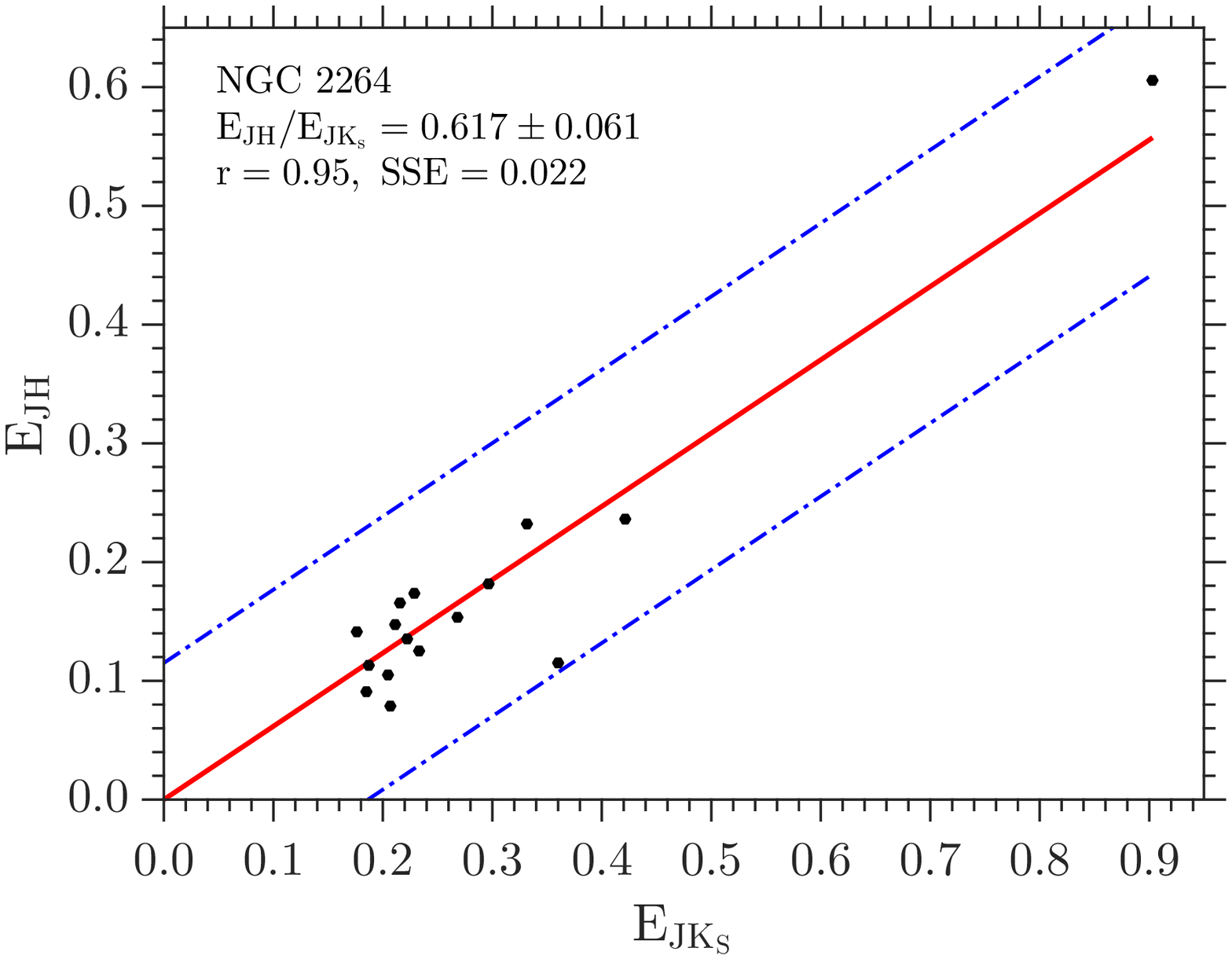}
   \includegraphics[scale=0.45]{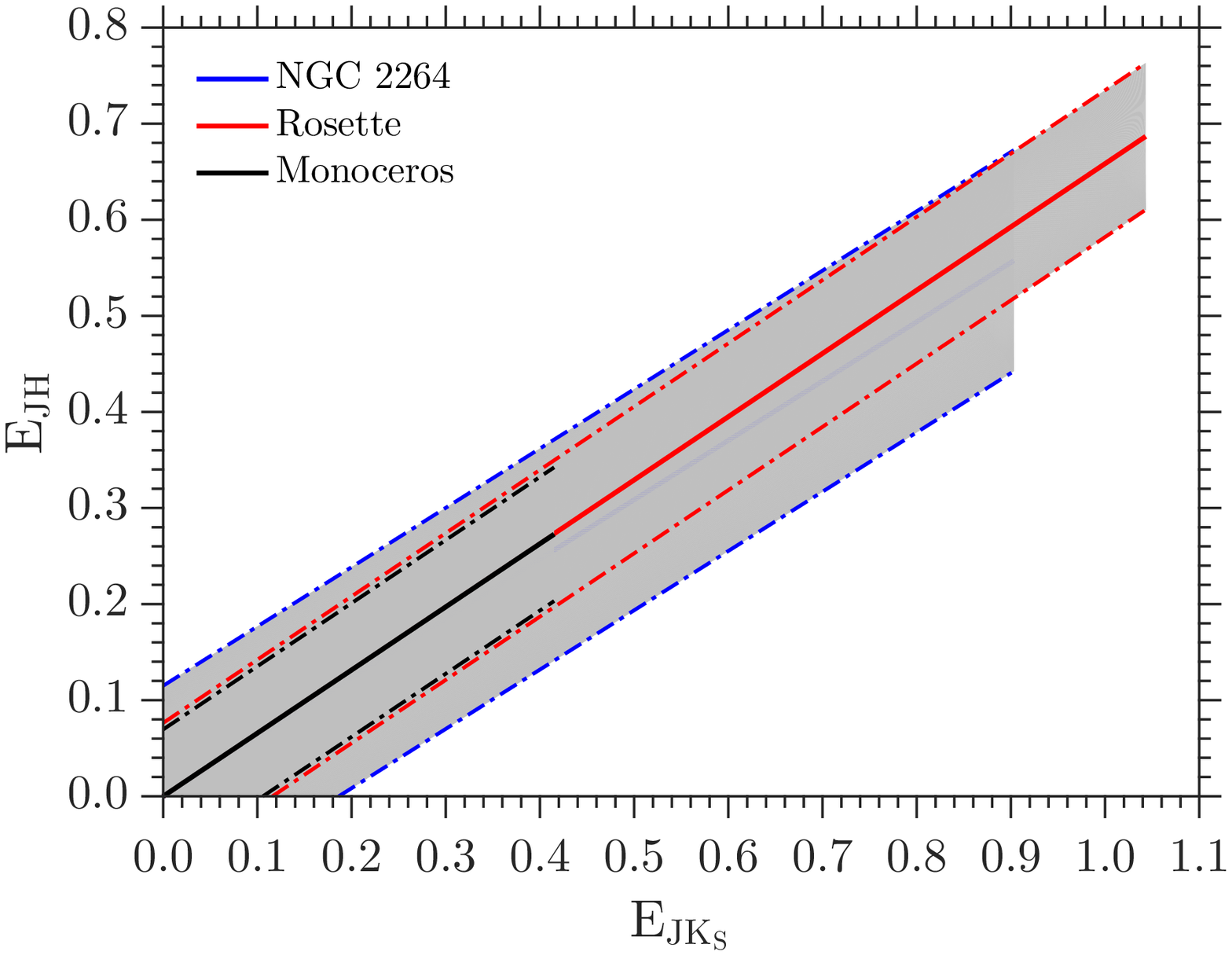}
   \caption{Color excess ratio, $\EJHJKs$, for the three target nebulas and their
   comparison. The red solid line is the linear fitting result, and the blue dash
   lines bound the 3$\sigma$ uncertainty region.
   \label{ceraneb}}
\end{figure*}

Stars behind the nebula are obscured by dust both from the nebula and the diffuse 
foreground ISD. But the nebula is inhomogeneous, they experience different extent 
of extinction by the nebula. The extinction by the nebula is calculated by subtracting 
the interstellar foreground extinction. With the nebular distance derived above, 
the stars further than this distance are chosen to study the extinction law of the 
nebula. More over, only the stars with apparent extinction by the nebula are taken 
as the tracers. In Figure \ref{avdmon}, the red dots with errorbars denote the 
extinction tracers that lie above the 3-sigma level of the background extinction 
and are used as the tracer stars of the nebular extinction. The same is for the 
Rosette nebula and NGC 2264, as shown in Figures \ref{avdros} and \ref{avdngc}. 
After subtracting the contribution by the background ISM, the color excess ratio, 
$\EJHJKs$, is derived by a linear fitting between $\EJH$ and $\EJKs$ as shown in 
Figures \ref{ceraneb} and in Table \ref{ceraall}.

The color excess ratio $\EJHJKs$ is $0.657 \pm 0.056$ for the Monoceros SNR,
$0.658 \pm 0.018$ for the Rosette nebula, which agree with each other, and also
with 0.652 by \citet{xue16}. As Monoceros is an old faint SNR, $\EJH$ and $\EJKs$
span a narrow range, which leads to a relatively large uncertainty (0.056) and 
low correlation coefficient ($r=0.89$). NGC 2264 has a smaller ratio, $\EJHJKs=
0.617$, but with an error of 0.061, it is still consistent with the mean value 
0.652. \citet{WJ14} suggest that the near-infrared extinction law is universal 
based on the fact that there is no visible change of $\EJHJKs$ with $\EJKs $ in 
the range $[0.3,~4.0]$. The Monoceros SNR shows no significant difference in the 
near-infrared extinction law from the mean law of the Milky Way, which conforms 
the universality of the near-infrared extinction law. However, the supernova 
explosion is a very violent event that releases numerous high energy particles 
and photons which can destroy the surrounding dust grains. Moreover, the supernova 
ejecta produce dust grains that may differ from the dust in the diffuse medium. 
In principle, the properties of the SN dust are expected to differ so is the 
extinction law. The highly consistency of the near-infrared extinction law of the 
two environments does not necessarily mean the SN dust is the same as others or 
the SN explosion has no effect on the surrounding dust grains. One possibility 
is that the Monoceros SNR is so old ($10^5$\,yr) that the dust observed is almost 
the normal ISD with little affected by the SN explosion. The other possibility 
is that the near-infrared bands cannot trace the difference of the dust. The other 
bands, in particular the visual and UV bands, may better reflect the difference of 
the dust.

\begin{table}
\begin{center}
\caption{\label{ceraall} Color excess ratio $\EJHJKs$ of the three nebulas.}
\begin{tabular}{lcccc}
\tableline \tableline
          & Monoceros SNR & Rosette Nebula & NGC 2264 & \citet{xue16} \\
$\EJHJKs$ & 0.657         & 0.658          & 0.617    & 0.652         \\
\tableline
\end{tabular}
\end{center}
\end{table}

\section{Dust Mass of the Monoceros SNR} \label{dustmass}

In principle, the dust mass of the Monoceros SNR can be derived from its extinction
because the extinction is proportional to the dust column density. A precise
determination of the dust mass needs the information of the extinction at all
wavelengths from which the dust property can be precisely constrained. Nevertheless,
a rough estimation of the dust mass can still be derived with the extinction known
only in the near-infrared if an extinction law is assumed.

Adopting the WD01 \citep{WD01} dust model for the Galactic interstellar 
extinction law ($R_{\rm V}=3.1$), the mass extinction coefficient for the $V-$band,
$K_{\rm ext,V} = \AV / \Sigma_{\rm dust}$ is

\begin{equation}
K_{\rm ext,V}=2.8 \times 10^4 {\rm ~mag ~cm^2 ~g^{-1}}.
\end{equation}

With a surface mass density $\Sigma_{\rm dust} = \AV/K_{\rm ext,V}$, the dust
mass is then

\begin{eqnarray}
\nonumber M_{\rm dust}
&& = \Sigma_{\rm dust} \times A_{\rm eff} \\
&& = \frac{\AV}{K_{\rm ext,V}} \times A_{\rm eff},
\end{eqnarray}

\noindent where $A_{\rm eff}$ is the effective surface area.

As a test of this method, we firstly apply it to the SN dust in the Crab 
Nebula which appears to be a $4.0 \times 2.9$\,pc ellipsoid \citep{hes08}. \citet{OB15} 
presented a detailed description of the nebular geometry. To calculate $A_{\rm eff}$, 
we follow the dust distribution of their favored models (\uppercase\expandafter{\romannumeral5} 
and \uppercase\expandafter{\romannumeral6}): a clumped shell starts at inner axis 
diameters of $2.3 \times 1.7$\,pc, and extends to the $4.0 \times 2.9$\,pc outer 
boundaries, with a volume filling factor ($\Ffil$) of 0.10. If we adopt $\AV = 
1.6 \pm 0.2$\,mag derived by \citet{mil73}, the resultant dust mass is $0.658 \pm 
0.082$\,$\Msun$ (the uncertainty is simply derived by using $\Delta \AV = 0.2$). 
This value is in agreement with that by \citet{OB15} who yielded a result of $0.11
-0.13$\,$\Msun$ of amorphous carbon and $0.39-0.47$\,$\Msun$ of silicate from the 
infrared emission by using mixed dust chemistry model. However, assuming a single 
dust species of carbon grains, \citet{gom12} derived warmer ($64 \pm 4$\,K) and 
cooler ($34 \pm 2$\,K) components of $0.006 \pm 0.02$ and $0.11 \pm 0.02$\,$\Msun$, 
respectively, and \citet{OB15} derived $0.18-0.27$\,$\Msun$ of amorphous carbon 
from clumped models. Both results are lower than our estimate. The discrepancy 
may be attributed to the value of $K_{\rm ext,V}$ which is affected by the species 
and size distribution of dust grains. \citet{noz13} construct a graphite-silicate 
model with a power law size distribution, which is similar to the mixed models of 
\citet{OB15}, and obtain $K_{\rm ext,V}=(3.7 \pm 0.5) \times 10^4$\,$\rm mag~cm^2~
g^{-1}$, which would make our estimation of dust mass being 0.498\,$\Msun$ and 
effectively reduce the discrepancy.

According to the distribution of the nebular tracers, a similar clumped-shell 
geometry as described by \citet{OB15} can be applied to Monoceros SNR. The SN 
explosion cleared an inner region around the central point so it is free of dust 
now, whilst the ISD has been swept-up into the outer dense shell, i.e. the clumped 
shell. Figure \ref{triDis}\,(a) shows the lack of significant extinction in the 
central part of the SNR, consistent with the presumed scenario. The Monoceros SNR 
has an angular diameter of 220\arcmin, corresponding to a radius of 63.36\,pc at 
the derived distance of 1.98\,kpc. We assume a circular shell for simplicity. The 
dust clumps start at inner radius, $\Rin$. From Figure \ref{avdmon}, it can be 
seen that the nebular extinction varies from about 0.01 to 0.15 in $\AKs$. For a 
rudimentary estimation, an average extinction of 0.05 in the $\Ks-$band is adopted 
that corresponds to 0.5\,mag in $\AV$. Then the mass of the dust ($\Mdust$) clumped 
in the shell is

\begin{eqnarray}
\nonumber \Mdust
&& = \frac{0.5 \times \pi \times (\Rout^2 - \Rin^2) \times \Ffil}{K_{\rm ext,V}} \\
&& = \left(1073.595 - 0.26743 {\left(\frac{\Rin}{\rm pc}\right)}^2 \right) \Ffil ~\Msun.
\end{eqnarray}

Because our extinction map is incomplete for the SNR due to the lack of
data, it is hard to determine the boundary of the inner ring. If the filling factor 
$\Ffil$ equals to 0.1 as \citet{bar10}, the dust mass is from 38.65\,$\Msun$ to 
80.52\,$\Msun$ if $\Rin$ is 50\% to 80\% of $\Rout$ estimated from Figure \ref{triDis}\,(a). 
Since the supernova dust is usually on the order of a few percent to at most a 
couple of tenths solar mass, the dust mass is mostly contributed by normal ISD. 
This fact can be understood by the old age of the Monoceros supernova remnant 
able to sweep a large region of ISM. This result is also consistent with the 
fact that the near-infrared extinction law agrees with the mean law as discussed 
in previous section. In this case, the characteristics inhibited in the SN explosion 
is obliterated when the ISD dominates absolutely during the long evolution after 
explosion. This method can be improved by an extinction law covering a complete 
wavelength range instead of only the $V-$band. We will modify the method in further 
work.

\section{Summary} \label{sum}

The goal of this work is to investigate the dust property of the SNRs from the 
nebular extinction and its law. The present work determines the distance and 
near-infrared extinction law of the Monoceros SNR and its nearby two nebulae
-- the Rosette Nebula and NGC 2264. By taking the stars in the corresponding
sightlines as the extinction tracers, the distance of a nebula is found at the
position of sharp increase of stellar extinction with distance. The stellar
extinction is calculated by its color excess with the intrinsic color index
derived from its stellar parameters (mainly $\Teff$) based on spectroscopic 
surveys. Its distance is calculated from the absolute magnitude fitted by the 
PARSEC model from $\feh$, $\Teff$ and $\logg$ after subtracting interstellar 
extinction. The distance of Monoceros SNR is 1.98\,kpc, larger than previous 
results. The distance of Rosette Nebula, 1.55\,kpc, agrees with some of previous 
values. The large difference between these two nebulae, 0.4\,kpc, implies little 
possibility that they are interacting with each other. For NGC 2264, the distance, 
1.2\,kpc, is slightly larger than previous results. The relative position of the 
three nebulae coincides with the \citet{dav78} result, i.e. the Monoceros SNR 
being the furthest and NGC 2264 the closest. The nebular extinction is derived by
subtracting the foreground extinction which is calculated from a reference diffuse
field with comparable Galactic latitude. The near-infrared extinction law of the
Monoceros SNR as well as the two nearby nebulas shows no apparent difference
with the mean near-infrared extinction law. This fact may be a piece of evidence
for the universality of the near-infrared extinction law. On the other hand, the
old age ($\sim$$10^5$\,yr) and the large mass ($\sim$50\,$\Msun$ on average) of 
Monoceros SNR signify that the material of this SNR is absolutely dominated by 
the ISD other than the SN ejecta. The work needs to be extended to the UV/visual 
extinction law and a more accurate estimation of the property of the SNRs.

\begin{acknowledgements}
We thank Profs. Bruce Draine, Jian Gao, Aigen Li, Yong Zhang, and the anonymous 
referee for very helpful suggestions and stimulating comments. We thank Mengfei 
Zhang for technological help. This work is supported by NSFC through Projects 
11373015, 11533002, and 973 Program 2014CB845702. This work makes use of the 
data from the surveys by LAMOST, SDSS/APOGEE, and 2MASS.
\end{acknowledgements}

\software{PARSEC, CMD (v3.0; http://stev.oapd.inaf.it/cmd)}

\bibliographystyle{aasjournal}
\bibliography{bibfile}

\end{document}